\documentclass[article,nofootinbib]{revtex4-1}

\usepackage{float}
\usepackage{epsfig}
\usepackage{dsfont}
\usepackage{amsmath}
\usepackage{braket}
\usepackage{amsfonts}
\usepackage{amssymb}
\usepackage{graphicx}  
\usepackage[dvipsnames]{xcolor}
\usepackage{todonotes}
\usepackage{bbm}
\usepackage{tikz}
\usepackage{lmodern}
\usepackage{braket}
\usepackage{wrapfig}
\usepackage{comment}
\usepackage{xcolor}
\usepackage{multirow}
\usepackage{soul}
\usetikzlibrary{decorations.pathmorphing}
\tikzset{snake it/.style={decorate, decoration=snake}}
\usepackage{fancybox}
\expandafter\ifx\csname package@font\endcsname\relax\else
\expandafter\expandafter
\expandafter\usepackage
\expandafter\expandafter
\expandafter{\csname package@font\endcsname}%
\fi

\RequirePackage[colorlinks,citecolor=blue,urlcolor=magenta,linkcolor=blue]{hyperref}

\newcommand{\be}{\begin{equation}}
\newcommand{\ee}{\end{equation}}
\newcommand{\bea}{\begin{eqnarray}}
\newcommand{\eea}{\end{eqnarray}}

\begin{document}
	
	\title{Infrared signatures of quantum bounce in a minisuperspace analysis of Lema\^{\i}tre-Tolman-Bondi dust collapse}

	\author{Harkirat Singh Sahota}
	\email{ph17078@iisermohali.ac.in }
	\author{Kinjalk Lochan}
	\email{kinjalk@iisermohali.ac.in}
	\affiliation{Department of Physical Sciences, Indian Institute of Science Education \& Research (IISER) Mohali, Sector 81 SAS Nagar, Manauli PO 140306 Punjab India.}

	\begin{abstract}
		Mode decomposition is crucial for studying the dynamics of propagation in a quantum process. In the quantum treatment of collapsing matter, a viable mode analysis is supposed to give information regarding emission during the collapse. Nevertheless, partly owing to operator ordering ambiguities involved in a typical quantum gravity analysis, the availability of such well-defined modes is not guaranteed. We study the mode decomposition of the unitarily evolving wave packet constructed for the quantum model of spherically symmetric dust collapsing in a marginally bound Lema\^{\i}tre-Tolman-Bondi (LTB) model. We consider a minisuperspace model of dust collapse, where black hole singularity is replaced by a bounce from collapsing phase to expanding phase in the quantum dynamics of the dust cloud. We identify the observable depicting mode decomposition, and using the freedom of operator ordering ambiguity, we obtain the Hermitian extension of this operator alongside the Hermitian Hamiltonian. After identifying incoming and outgoing modes with this operator's eigenstates, we estimate their contributions to the radiation profile. True to a quantum description, the expanding and contracting branches do not entirely comprise of outgoing and incoming radiation. The infrared sector of this process demonstrates some characteristic features which turn out to be highly sensitive to the near-bounce dynamics of the dust cloud. Near the epoch of classical singularity, there is a significant contribution from incoming/outgoing modes of small wave number in the expanding/collapsing phase of the dust cloud, which keeps on decreasing as one moves away from the singularity. The information of the bounce is carried over to the infrared modes through a flip from largely incoming to largely outgoing radiation as the evolution progresses from collapsing to expanding phase, much before the information of bounce comes about to any observer. In the infrared sector, the saturation value of the amplitude marks the bounce radius. Thus, we argue that the information of the short scale physics is essentially carried over to the longest wavelength in this quantum gravity model, which we argue is rather more prominent for low energy processes.
	\end{abstract}
	\maketitle
	
	\section{Introduction}
	Successful quantization of the theory of gravity is one of the long-standing problems in theoretical physics. From the inception of this idea, it is expected that implementing the principles of quantum theory for the case of gravity might help one to better understand the issue of singularities that plagues the classical theory, e.g. black holes or the early universe. Even as early as the 1930s, attempts were being made to develop a consistent quantum theory of gravity \cite{rocci_first_2013}. However, a more systematic analysis of the problem started with the advent of the ADM formalism, i.e., Hamiltonian formulation of general relativity \cite{arnowitt_dynamical_1959}.
	
	General relativity by its structure,  is a singular theory, i.e., the system has some fiducial degrees of freedom and it is an example of a gauge (constrained) system with first class constraints \cite{henneaux_quantization_1992,prokhorov_hamiltonian_2011}. These constraints are the Hamiltonian constraint and diffeomorphism constraints which generate time reparametrizations and diffeomorphism transformations, the gauge transformations of a generally covariant theory. DeWitt implemented Dirac's criteria of quantizing singular systems and wrote down what now is called the Wheeler-DeWitt equation \cite{dewitt_quantum_1967,wheeler_superspace_nodate}. This approach to tackle the problem of quantization of gravity similar to a gauge theory is called canonical quantization of gravity \cite{kiefer_quantum_2012}. The scheme of canonical quantization of gravity is adopted in various avatars with different phase space variables to analyze the constraints more effectively \cite{ashtekar_new_1986,rovelli_loop_2008,bojowald_loop_2008}.  After identifying the canonical variables to quantize, the constraint equations are expressed in terms of the phase space variables of the system. A quantum state correctly describing a gravitational system must satisfy the Wheeler-DeWitt equation. Unfortunately, it has not been possible to get the solutions of the Wheeler-DeWitt equation in full generality in any of the canonical quantization schemes. Nonetheless, just as in the classical general relativity, the symmetry reduced models are easier to handle and are studied extensively in the canonical approach to quantization \cite{kiefer_quantum_2012,bojowald_quantum_2015}. In geometrodynamics, configuration space is in general an infinite-dimensional space of all three-metrics distinguishable up to a diffeomorphism. Symmetry reduction is used to slash down the degrees of freedom of the system, restricting the action to those which explicitly carry the symmetries from the beginning. In such symmetry reduced systems, if there still remain infinitely many degrees of freedom, the process is identified as a {\it midisuperspace reduction} (e.g. restricting the theory to spherically symmetric sector \cite{barbero_g_quantization_2010,torre_midisuperspace_1999}) and if the symmetry reduction is vast such that only  finitely many  degrees of freedom are left in the configuration space, the reduction is called {\it minisuperspace reduction} (e.g. homogeneous and isotropic universe with a single degree of freedom, the scale factor $a(t)$ \cite{kiefer_quantum_2012,kiefer2008quantum,wiltshire_introduction_2003,halliwell_introductory_2009,coule_quantum_2005}).
	
	Reducing the symmetry classically first and then quantizing the theory may appear appealing from the vantage point of potential simplification of constraint equations. However, the conceptual problem with such a reduction before quantization is that it may violate the uncertainty principle of the full theory, because one may freeze both coordinates and their conjugate momenta \cite{kuchar_is_1989,bojowald_minisuperspace_2016} (quantizable variables of the full theory).
	Also, it will be imperative to check for anomalies in such models \cite{kiefer_quantum_2012,bojowald_minisuperspace_2016}. We need to be careful because it might be the case that the symmetries we hold dear might be broken in the true quantum model.
	Therefore, these minisuperspace/midisuperspace models must be considered as toy models, which one hopes, will nonetheless capture some essence of the full quantum gravity theory. For example, in umpteen numbers of mini/midisuperspace quantization studies, the fate of classical singularities has remained  a primary research goal \cite{Dabrowski:2006dd,bergeron_singularity_2015,liu_singularity_2014,wilson-ewing_loop_2018,kamenshchik_quantum_2013,sami_avoidance_2006,ashtekar_quantum_2006,ashtekar_quantum_2006-1,ashtekar_loop_2009,Gielen:2020abd,Gielen:2021igw,ashtekar_loop_2009-1,hajicek_singularity_2001,ambrus_quantum_2005,vaz_canonical_2011,kiefer_quantum_2015}. 
	
	The quantization of collapsing dust shell and singularity resolution in the context of quantum geometrodynamics is addressed in \cite{hajicek_embedding_2001,hajicek_singularity_2001,ambrus_quantum_2005,vaz_canonical_2011,kiefer_quantum_2015}. The demand of unitary evolution in this model \cite{hajicek_singularity_2001} leads to vanishing of the wave function at the time of classical singularity, and thus singularity is avoided according to DeWitt criteria \cite{dewitt_quantum_1967}. The dynamics of the dust shell is represented by the wave packet constructed in this model, which shows that the shell collapses past the horizon to a minimal radius and then starts reexpanding. Quantization of a collapsing dust shell also has been studied in the context of loop quantum gravity \cite{rovelli_planck_2014,ashtekar_black_2005,bambi_terminating_2014,Giesel:2021dug}. Again, the collapse of the dust shell to singularity is replaced by the bounce of quantum shell from collapsing branch to expanding branch. The lifetime of a black hole-like temporary object and behavior of the horizon is a major issue in this approach. For these models to be realistic, this lifetime should be greater than the current age of the Universe, as one would like to apply these models to describe the quantum collapse of astrophysical objects and relate to possible observations. See \cite{malafarina_classical_2017} for a recent review.
	
	A midisuperspace quantization scheme to tackle black hole singularity for the case of dust collapse in the LTB model is developed in \cite{vaz_toward_2001}. Over many years, various aspects of this model such as Hawking radiation, nonthermal corrections and the associated entropy have been investigated \cite{vaz_quantum_2001,Vaz:2001mb,vaz_quantum_2003,kiefer_classical_2006,kiefer_hawking_2007,vaz_quantum_2007,Franzen:2009ev,banerjee_quantum_2010,Vaz:2011zz,vaz_tunneling_2013}. In this approach, Hawking radiation is envisaged as the projection of a wave functional along the outgoing plane wave basis. In order to keep the process unitary, a suitable measure is chosen to keep the Hamiltonian Hermitian, which together with the kernel space, gives the Hilbert space of quantum gravity.
	
	To study the mode decomposition in the quantum model, one needs to identify an observable in the phase space, which closely follows the classical incoming/outgoing character of the dust cloud. The eigenstates of such an observable will be well suited as incoming and outgoing modes. However, since the canonical momentum does not typically commute with the Hamiltonian, it is not always possible in a general quantum gravity model to come up with a measure that simultaneously keeps the Hamiltonian and the momentum Hermitian. To work with an arbitrary measure, the conjugate momenta should also assume some valid representation to maintain Hermiticity, if possible. In the absence of that, the notion of the incoming and outgoing wave remains mathematically somewhat ill defined, subsequently challenging the quantum analysis of emission spectra from a collapsing cloud using Bogoliubov coefficients \cite{vaz_quantum_2007}. 
	
	In this work, we study the interplay of the measure with the representation of the momentum operator supporting well-defined orthogonal ingoing and outgoing states. For this purpose, we start with the minisuperspace formalism of dust collapse in a marginally bound LTB model  \cite{kiefer_singularity_2019}. Using the fact that different dust shells are decoupled, in this model, one can obtain the effective action which gives the dynamics of the outermost shell, from which one can infer the behavior of the full dust cloud. Since dust naturally provides a preferred notion of time, a quantum realization of this effective minisuperspace model is developed using Brown-Kucha\v{r}'s construction \cite{brown_dust_1995}. In this quantum model, the classical singularity is shown to be replaced by a bounce followed by expansion.

	The Hamiltonian of this system suffers from operator ordering ambiguity in its quantum avatar. Given a particular operator ordering, one has to select the measure judiciously in order to make the Hamiltonian Hermitian, ensuring unitary time evolution.  This work is motivated by the observation that, in addition to the Hamiltonian, one can make the momentum conjugate to the shell area-radius a Hermitian operator in this model, for a certain choice of operator ordering parameters or the representation of momentum operator. 
	
	In terms of the eigenstates of the momentum, we study the mode decomposition of the unitarily evolving wave packet constructed in \cite{kiefer_singularity_2019}. For a well-defined quantum state, the expanding (contracting) branches of the dust shell dynamics do not entirely comprise of outgoing (incoming) modes but contain a certain fraction of incoming (outgoing) radiation too. The infrared sector of this process shows some interesting features. Near-infrared modes of the wave packet are shown to be very sensitive to the dynamics. If one focuses on the infrared sector, the information of bounce is carried over to these modes, much before the classical information of the bounce arrives otherwise \cite{kiefer_singularity_2019}. The infrared sector quickly adopts the characteristic of the dynamics, leading to a flip from a largely incoming character to a largely outgoing one, as the evolution progresses from collapsing to expanding branch. Further, the emission amplitude saturates to a value in the far infrared regime, which is directly proportional to the bounce radius. Thus information of short scale physics is effectively carried over to the longest wavelength in such quantum gravity models. Moreover, comparing the bounce radius with the energy of the initial dust cloud, it can be shown that such infrared effects become more significant for low energy collapse, somewhat counter-intuitively.
	
	This paper is organized as follows. In Sec. \ref{MarginalLTB}, we review the marginal LTB dust collapse model. We summarize the findings of the quantum model constructed in \cite{kiefer_singularity_2019} in this section. In Sec. \ref{Mode decomposition}, we study the decomposition of the wave packet into ingoing and outgoing modes for the case when the measure chosen is $R^2$. In Sec. \ref{Infrared} we study the infrared behavior of the collapse process in terms of the modes obtained from the momentum operator. We show how the information of the bounce and bounce radius are carried over to the infrared sector. In Sec. \ref{Rgen} we briefly extend our analysis when a general measure is chosen as $R^{1-a-2b}$. In Sec. \ref{Obsdep}, we show that the expectation value of any general phase space observable for any general wave packet constructed is independent of operator ordering parameter b and depends only on a. We conclude our findings in Sec. \ref{Consclusions}.

	\section{On-Shell construction of Marginal LTB model} \label{MarginalLTB}
	The LTB model is an inhomogeneous extension of the Friedmann-Robertson-Walker model, which has spherical symmetry and a nonrotational dust of energy density $\epsilon$ acting as its source \cite{Krasinski:1997yxj}. The system is represented by the line element and Einstein equations given respectively as,
	\begin{align}
	ds^2=-c^2d\tau^2+\frac{R'^2}{1+2f(\rho)}d\rho^2+R^2d\Omega^2,\\
	\frac{F'}{R^2R'}=\frac{8\pi G\epsilon}{c^2}\quad \text{and}\quad \frac{R\dot{R}^2}{c^2}=F+2fR,\label{eom}
	\end{align}
	where $R(\tau,\rho)$ is the areal/curvature radius of the shell labeled by $\rho$ at time $\tau$, and $f(\rho)$ is called the energy function. The function $F(\rho)$ is equal to twice that of the Misner-Sharp(MS) mass \cite{szabados_quasi-local_2009} for LTB spacetime, $M_{MS}=R\dot{R}^2/2-fR=F/2$.  In spherically symmetric gravitational collapse, the Misner-Sharp mass is interpreted as the total gravitating mass enclosed inside a dust shell located at $\rho$. In this analysis, we will work with the marginally bound LTB model, for which $f(\rho)=0$.	
	
	The object of interest in this model is $R(\tau,\rho)$, which describes the dynamics of the dust cloud. This model has a singularity when the dust cloud collapses to a point, i.e., at $R=0$. The equation of motion dictating the dynamics of $R$ given by Eq. \eqref{eom} depends only on $R$ and $F$, but not on their spatial derivatives, which implies that for a given mass function $F$, the different dust shells are dynamically decoupled. Thus the different dust shells can be considered independently, and we can quantize the outermost shell in a marginally bound LTB model. The dynamics of the full dust cloud is then deduced from this model \cite{kiefer_singularity_2019}.

	To derive the action of the outermost dust shell, we start from the Einstein-Hilbert action
	\begin{align}
	S=\frac{1}{16\pi}\int_\mathcal{M}d^4x\sqrt{-g}\mathcal{R}[g]+\frac{1}{8\pi}\int_{\partial\mathcal{M}}d^3x\eta\sqrt{|h|}(\text{k}-\text{k}^0),
	\end{align}
	where k is the trace of extrinsic curvature of boundary $\partial\mathcal{M}$ and $\text{k}^0$ is its value when the boundary is embedded in flat space. For timelike hypersurface, $\eta$ is 1 and for spacelike hypersurface, it is $-1$. Using Einstein's equations, Ricci scalar takes the form $\sqrt{-g}\mathcal{R}[g]= 8\pi\epsilon\;R^2R'\sin{\theta}=F'\sin{\theta}$. Integrating out angular coordinates and using this expression for the on-shell Ricci scalar, the bulk part of Einstein-Hilbert action $S_{\mathcal{M}}$ takes the form
	\begin{align}
	S_{\mathcal{M}}=\frac{1}{4}\int d\tau \int_0^{\rho_o}d\rho\;F'=\frac{1}{4}\int d\tau\;F_o=\frac{1}{4}\int d\tau\;R_o\dot{R}_o^2.
	\end{align}
	
	Here $R_o$ is the curvature radius at $\rho_o$ the location of the outermost dust shell. In further analysis, we will remove the subscript $o$ and $R$ will be the curvature radius associated with the outermost dust shell. The boundary $\partial\mathcal{M}$ comprises of the union of two spacelike hypersurfaces of fixed constant dust proper time $\tau_1\;\text{and}\; \tau_2,\text{ with }\; \tau_1<\tau_2$, and a timelike hypersurface coinciding with the outermost dust shell at $\rho=\rho_o$. We can write the GHY(Gibbons-Hawking-York) term for the boundary hypersurfaces, which receives contribution only from spacelike hypersurfaces as for timelike hypersurface $\text{k}$ and $\text{k}^0$ are equal. The boundary contribution to the action from these spacelike hypersurfaces is \cite{kiefer_singularity_2019},
	\begin{align}
	S_{\partial\mathcal{M}}=-\frac{3}{4}\int d\tau R\dot{R}^2.
	\end{align}
	Adding that to the bulk part, we can write the action which dictates the dynamics of the outermost shell,
	\begin{equation}
	\mathcal{S}=-\frac{1}{2}\int d\tau R\dot{R}^2\label{a}.
	\end{equation}
	As the equations of motion are used to reach at this action, we are left with a prescription for how boundary at the initial time evolves into the future. The Hamiltonian associated with this action is,
	\begin{equation}
	H=-\frac{P^2}{2R}=-E_{ADM}.
	\end{equation}
	This Hamiltonian is the negative of the ADM energy. In \cite{kiefer_singularity_2019}, the Brown-Kucha\v{r}'s prescription for dust \cite{brown_dust_1995} is used as the matter and its proper time as the time coordinate, which is a standard prescription to deal with the problem of time in quantum gravity \cite{vaz_toward_2001,vaz_quantum_2003,Montani:2003qq,Battisti_2006,Montani:2007vu,Amemiya_2009,Giesel_2010,Husain:2011tm,maeda_unitary_2015,kiefer_singularity_2019}. The Hamiltonian constraint for this model takes the form,
	\begin{equation}
	\mathcal{H}\equiv p_\tau+H\approx0.
	\end{equation}
	Classically, in this minisuperspace model, the dynamics of dust shell is dictated by the equation of motion,
	\begin{align}
	R\dot{R}^2=2E
	\end{align}
	Here $E$ is ADM energy. The solution is given by the equation,
	\begin{align}
	R(\tau)=\left(\frac{3}{2}\sqrt{2E}|\tau|\right)^{\frac{2}{3}}.\label{A5}
	\end{align}
	It represents two classically disjointed branches, a collapse from $\tau=-\infty$ to $\tau=0$ and an expansion from $\tau=0$ to  $\tau=\infty$. Classical singularity is the endpoint of collapse (a black hole singularity) when the dust shell collapses to $R=0$ at $\tau=0$ or starting point of the expansion (a white hole singularity) $R=0$ at $\tau=0$.
	
	Following Dirac's prescription of quantizing constraint systems, we can write the Wheeler-DeWitt equation for this model as
	\begin{align}
	i\hbar\frac{\partial \Psi(R,\tau)}{\partial\tau}&=\hat{H}\Psi(R,\tau),\\
	\hat{H}=\frac{\hbar^2}{2}R&^{-1+a+b}\frac{d}{dR}R^{-a}\frac{d}{dR}R^{-b}.\label{Hamlitonian}
	\end{align}
	Since the classical Hamiltonian involves a product of $R$ and $P$, it does not have a unique quantum counterpart. Therefore, this model exhibits the operator ordering ambiguity. The parameters $a$ and $b$ in Eq. \eqref{Hamlitonian} represent our freedom to choose operator ordering. Using separation ansatz, we can solve the time-independent equation $\hat{H}\phi_E=-E\phi_E$. With $\hbar=1$ the stationary states are \cite{kiefer_singularity_2019},
	\begin{align}
	\begin{aligned}
	\phi_E^1(R)&=R^{\frac{1}{2}(1+a+2b)}J_{\frac{1}{3}|1+a|}\left(\frac{2}{3}\sqrt{2E}R^{\frac{3}{2}}\right),\\
	\phi_{-E}^1(R)&=R^{\frac{1}{2}(1+a+2b)}I_{\frac{1}{3}|1+a|}\left(\frac{2}{3}\sqrt{2E}R^{\frac{3}{2}}\right),\\
	\end{aligned}\\
	\begin{aligned}
	\phi_E^2(R)&=R^{\frac{1}{2}(1+a+2b)}Y_{\frac{1}{3}|1+a|}\left(\frac{2}{3}\sqrt{2E}R^{\frac{3}{2}}\right),\\
	\phi_{-E}^2(R)&=R^{\frac{1}{2}(1+a+2b)}K_{\frac{1}{3}|1+a|}\left(\frac{2}{3}\sqrt{2E}R^{\frac{3}{2}}\right),
	\end{aligned}\\
	\phi_0^1=R^b\quad,\quad\phi_0^2=R^{1+a+b},\qquad\qquad&
	\end{align}
	where $J_n,\;Y_n,\;K_n\;\;\text{and}\;I_n$ are Bessel's functions of the first and second kind. Here $E$ can be interpreted as the ADM energy $E_{ADM}$. Classically the ADM energy $E_{ADM}=R\dot{R}^2/2$ is always positive, but the quantum Hamiltonian operator has positive eigenvalues (negative ADM energy) as well in its spectrum, which can be interpreted as genuine quantum solutions without any classical counterpart.  We choose Hilbert space with inner product $L^2(\mathbb{R}^+,R^{1-a-2b}dR)$ that will make this Hamiltonian Hermitian,
	\begin{align}
	\braket{\psi|\chi}=\int_{0}^{\infty}dR\;R^{1-a-2b}\psi^*(R)\chi(R).\label{inner}
	\end{align}
	The self-adjoint extensions of the Hamiltonian \eqref{Hamlitonian} are discussed in \cite{kiefer_singularity_2019}. Since, $J_\nu$ functions have a closure relation, i.e.
	\begin{align}
	\int_{0}^{\infty}dx\;x\;J_\nu(\alpha x)J_\nu(\alpha' x)=\frac{\delta(\alpha-\alpha')}{\alpha},\;\text{for} \;\nu>-\frac{1}{2}
	\end{align}
	the positive energy stationary states $\phi_{E}^1$ form an orthogonal set under the chosen scalar product, thus making them suitable for the construction of a wave packet. 
	\begin{align}
	\braket{\tilde{\phi}_{E}^1|\tilde{\phi}_{E'}^1}=\delta\left(\sqrt{E}-\sqrt{E'}\right),\quad \;\text{where}\; \tilde{\phi}_{E}^1=\frac{2}{\sqrt{3}}E^{\frac{1}{4}}\phi_{E}^1, \quad \text{for }E>0.
	\end{align}
	From the positive energy modes, a unitarily evolving wave packet is constructed by choosing a normalized Poisson-like distribution \cite{kiefer_singularity_2019}
	\begin{align}
	\psi(R,\tau)&=\int_{0}^{\infty}d\sqrt{E}\tilde{\phi}_E(R)e^{i E\tau}A(\sqrt{E}),\\
	A(\sqrt{E})&=\frac{\sqrt{2}\lambda^{\frac{1}{2}(\kappa+1)}}{\sqrt{\Gamma(\kappa+1
			)}}\sqrt{E}^{\kappa+\frac{1}{2}}e^{-\frac{\lambda}{2}E},\label{A4}
	\end{align}
	where $\kappa\geq0$ and $\lambda>0$ are real parameters with $\kappa$ being dimensionless and $\lambda$ has dimensions of inverse of energy. For this choice of distribution, the expectation value of the Hamiltonian is
	\begin{align}
	\bar{E}=\braket{\psi|\hat{H}|\psi}=\frac{\kappa+1}{\lambda},\label{energy}
	\end{align}
	which is inversely proportional to $\lambda$. With this distribution the wave packet takes the form,
	\begin{align}
	\begin{aligned}
	\psi(R,\tau)=\sqrt{3}\left(\frac{\sqrt{2}}{3}\right)^{\frac{1}{3}|1+a|+1}&\frac{\Gamma\left(\frac{1}{6}|1+a|+\frac{\kappa}{2}+1\right)}{\sqrt{\Gamma(\kappa+1)}\Gamma\left(\frac{1}{3}|1+a|+1\right)}\frac{\lambda^{\frac{1}{2}(\kappa+1)}}{\left(\frac{\lambda}{2}-i\tau\right)^{\frac{1}{6}|1+a|+\frac{\kappa}{2}+1}}R^{\frac{1}{2}(1+a+|1+a|+2b)}\\
	&\;_1F_1\left(\frac{1}{6}|1+a|+\frac{\kappa}{2}+1;\frac{1}{3}|1+a|+1;-\frac{2R^3}{9\left(\frac{\lambda}{2}-i\tau\right)}\right).\label{gwp}
	\end{aligned}
	\end{align}
	To simplify the expression, a prescription $\kappa=|1+a|/3$ is adopted in \cite{kiefer_singularity_2019} and the expression for the wave packet reduces to, 
	\begin{align}
	\psi(R,\tau)=\sqrt{3}\frac{R^{\frac{1}{2}(1+a+|1+a|+2b)}}{\sqrt{\Gamma(\frac{1}{3}|1+a|+1)}}\left(\frac{\frac{\sqrt{2\lambda}}{3}}{\frac{\lambda}{2}-i\tau}\right)^{\frac{1}{3}|1+a|+1}e^{-\frac{2R^3}{9\left(\frac{\lambda}{2}-i\tau\right)}}\label{wp}.
	\end{align}
	This simplification makes the computation of various expectation values easier, but it comes at the cost of making the distribution a function of the operator ordering parameter. For example, to have well-defined energy, we would need to have a small value of the relative width, which in turn would require a very large value of parameter $a$. We have to be careful while interpreting the results because the signature of parameter $a$ in an observable can be either an artifact of operator ordering ambiguity or it can also signify its dependence on the shape of the distribution.

	Following DeWitt's criteria, the singularity is avoided if the probability amplitude vanishes at the location of the anticipated singularity in the classical model. We are interested in the behavior of the wave packet at $R=0$, which can be estimated by looking at the behavior of the stationary states at the singularity. For $z\rightarrow$ 0, at the leading order, $J_\nu$ behaves as
	\begin{align}
	J_\nu\sim\frac{1}{\Gamma(\nu+1)}\left(\frac{z}{2}\right)^\nu.
	\end{align}
	Hence the probability amplitude associated with $\phi_{E}^1$ behaves like,
	\begin{align}
	R^{1-a-2b}\phi_{E}^{1*}\phi_{E}^1\sim\left(\frac{2E}{9}\right)^{\frac{1}{3}|1+a|}\frac{R^{2+|1+a|}}{\Gamma(1+\frac{1}{3}|1+a|)^2}\rightarrow 0.
	\end{align}
	The probability amplitude vanishes for the stationary states and so do the wave packets constructed from these states. The expectation value of $R$ for this wave packet is given as,
	\begin{align}
	\bar{R}(\tau)=\braket{\psi|\hat{R}|\psi}=\frac{3\Gamma(\frac{1}{3}|1+a|+\frac{4}{3})}{2\sqrt{2}\Gamma(\frac{1}{3}|1+a|+1)}\left(\lambda+\frac{4\tau^2}{\lambda}\right)^\frac{1}{3},\label{A6}
	\end{align}
	$\bar{R}(\tau)$ is symmetric in $\tau$ and has global minimum at $\tau$=0, the minimal radius. For $\tau\in(-\infty,0)$, as time increases, $\bar{R}$ decreases, representing a contracting phase, while $\tau\in(0,\infty)$ represents the expanding phase.  Thus the classical collapse of dust to the singularity in the quantum model is replaced by the bounce. The expression of $\bar{R}(\tau)$ in \eqref{A6} diverges for $\lambda\rightarrow 0$, but that represents dust shell of infinite energy \eqref{energy}. The singularity is avoided for states representing dust shells with finite energy.
	
	The lifetime of the horizon (lifetime of gray hole state) can be estimated in this model \cite{kiefer_singularity_2019} which is interpreted as the time taken by the dust cloud to cross the apparent horizon twice. This lifetime from the point of view of comoving observer scales linearly with the mass of the dust cloud. The direct computation of the lifetime as observed by an outside observer would require transformation to Schwarzschild Killing time, which is ill defined in this model, although Schwarzchild exterior is incorporated for the dust collapse models in recent works \cite{PhysRevD.101.026016,kwidzinski_hamiltonian_2020,piechocki_quantum_2020,PhysRevD.103.064074}. Instead, a different approach is followed in \cite{kiefer_singularity_2019}, to compute the lifetime from transition between dynamically distinct states. The dynamics of the dust cloud is divided into three regimes, a collapsing regime when $\tau\leq-\tau_{AH}$, gray hole regime when $-\tau_{AH}\leq\tau\leq\tau_{AH}$ and expanding regime when $\tau\geq\tau_{AH}$, where $\tau_{AH}$ is the proper time at which the outermost shell reaches the apparent horizon. The lifetime is then defined as the time it takes for the dust cloud to go from a gray hole state to an expanding state, which comes out to be proportional to $\bar{E}^3$ \cite{kiefer_singularity_2019}. Thus to learn about the bounce, the outside observers have to wait for a long time for a sharply peaked distribution to arrive. We are interested in knowing, if the wave packet can be decomposed in another observable basis that is natural to quantify the incoming-outgoing modes. For this, the momentum operator is a good choice as classically, we have $P=-R\dot{R}$. Therefore we can associate positive values of momentum with collapsing phase and negative momentum with expanding phase of the dust cloud.

	
	\section{Mode decomposition of a wave packet}\label{Mode decomposition}
	The emission from the quantum collapse process potentially forming a black hole is typically obtained by estimating the contribution of outgoing part in the states specifying the collapsing dust cloud. To achieve that, incoming/outgoing modes are associated with the eigenstates of the Hamiltonian constraint operator.
	In \cite{vaz_quantum_2001,Vaz:2001mb,vaz_quantum_2003,kiefer_classical_2006,vaz_quantum_2007,kiefer_hawking_2007,Franzen:2009ev,banerjee_quantum_2010,Vaz:2011zz,vaz_tunneling_2013},  in the context of a midisuperspace construction, exact solutions of the Wheeler-DeWitt (WdW) equation are derived and quantum dynamics is deduced from them. The Hilbert space for such a model is defined with a measure that will make the Hamiltonian constraint Hermitian. After making the transformation from dust comoving time to Schwarzschild killing time, incoming/outgoing modes in these models are associated with the asymptotic limit of the solutions of Wheeler-DeWitt equation outside the dust cloud.\\
	In order to use the lattice regularization scheme for the midisuperspace collapse, a particular class of operator ordering is adopted  \cite{kiefer_classical_2006}. The Wheeler-DeWitt equation in this model is,
	\begin{align}
	\left[\frac{\delta^2}{\delta\tau^2}+\mathcal{F}\frac{\delta^2}{\delta R^2}+A\delta(0)\frac{\delta}{\delta R}+B\delta(0)^2-\frac{\Gamma^2}{\mathcal{F}}\right]\Psi[\tau,R,\Gamma]=0,
	\end{align}
	where $A$ and $B$ are smooth functions of the areal radius $R$ and the mass function $F$ that encapsulate the operator ordering ambiguity, $\mathcal{F}=1-F/R$ and $\Gamma=F'$ is the energy density. The formal expression $\delta(0)$ is included to indicate the need for the regularization. Working on latticized system is only possible provided the operator ordering term contributing to the potential term $B$ vanishes. In the regularized model, the inner product is introduced with a measure \cite{kiefer_hawking_2007},
	\begin{align}
	\braket{\Phi|\Psi}=\int_0^\infty dR_j\sqrt{g^{}_{RR}}\Phi^*(R_j)\Psi(R_j)\label{measure1}.
	\end{align}
	Here $g^{}_{RR}$ is the $RR$ component of the DeWitt metric and $R_j$ represents areal radius of the dust shell at $\rho_j$. With this measure though, both the Hamiltonian constraint and the momentum conjugate to $R$ are not Hermitian. The issue of Hermiticity of the Hamiltonian constraint and choice of measure that will make it Hermitian is discussed in \cite{vaz_quantum_2007}. The measure $\mu_j(R_j)$ in this case is given as
	\begin{align}
	\braket{\Phi|\Psi}=\int_0^\infty dR_j\mu_j(R_j)\Phi^*(R_j)\Psi(R_j).
	\end{align}
	The measure and operator ordering function are related via
	\begin{align}
		A_j=|\mathcal{F}_j|\partial_{R_j}\left[\log(\mu_j|\mathcal{F}_j|)\right].
	\end{align} 
	However, in this case as well, the outgoing/incoming modes are not necessarily orthogonal. We are interested in seeing whether it is possible to have both the momentum as well as the Hamiltonian constraint Hermitian in this model. Since a particular operator ordering of Hamiltonian is chosen to solve the Wheeler-DeWitt equation, the momentum operator is not Hermitian in the trivial representation,
	\begin{align}
	\braket{\Phi|\hat{P}|\Psi}=\int dR_j\mu_j(R_j)\Phi^*\frac{\delta\Psi}{\delta R_j}\neq\braket{\hat{P}\Phi|\Psi}.
	\end{align}
	In the absence of a momentum operator and modes associated with it, various prescriptions are used for defining incoming/outgoing modes. In \cite{vaz_quantum_2003}, a solution of the WdW equation outside the horizon is written in terms of the Killing time (obtained by matching the contracting dust cloud to the Schwarzschild exterior) and the limit $T\rightarrow-\infty$ and $R\rightarrow\infty$ of these wave functions give incoming modes. For outgoing modes, killing time is obtained by matching the expanding dust cloud to the Schwarzschild exterior and $T\rightarrow\infty$ and $R\rightarrow\infty$ limit of solution of the WdW equation written in terms of this Killing time. These modes take the plane wavelike form, providing a complete set of incoming/outgoing modes at each coordinate label $\rho$,
	\begin{align}
	\psi^-_\omega=\prod_{\rho}e^{-i \omega(\rho)[T(\rho)+Z(\rho)]},\text{ and }\psi^+_\omega=\prod_{\rho}e^{-i \omega(\rho)[T(\rho)-Z(\rho)]},\text{ with }Z(\rho)=4\sqrt{2MR}.
	\end{align}
	Whereas in \cite{kiefer_hawking_2007}, after making the transformation from dust comoving time to the Killing time, the incoming/outgoing modes are associated with the positive/negative frequency solutions of the WdW equation. The Bogoliubov coefficients in these models are then defined as the projection of outgoing mode functionals with the solution of WdW equation \cite{vaz_quantum_2003} and inner product of incoming and outgoing wave functionals \cite{kiefer_hawking_2007},
	\begin{align}
	\beta_{\omega\omega'}=\frac{2\sigma\omega}{G\hbar}\int_{R_h}^{\infty}dR\sqrt{g_{RR}}\Psi_{\omega}^{-*}\Psi_{\omega'}^+.
	\end{align}
	Although the Hawking radiation is recovered, the formalism of Bogoliubov coefficients is essentially reliant on the orthonormality of the mode functions \cite{Birrell:1982ix} whereas the incoming/outgoing modes introduced in these approaches are not orthonormal. This issue stems from the fact that the Hamiltonian is not Hermitian with the measure $\sqrt{g_{RR}}$ in these models. Even for the case when the Hamiltonian is Hermitian \cite{vaz_quantum_2007}, we do not expect the states to be orthogonal as they are degenerate states with zero eigenvalue. Therefore, the notion of incoming/outgoing modes and analysis of Bogoliubov coefficients remain somewhat ill defined. In the next subsection, we will argue that the freedom offered in selecting measure in the quantum LTB model which enables one to obtain an orthogonal set of incoming/outgoing modes.  In \cite{kiefer_singularity_2019}, the most general operator ordering for the Hamiltonian constraint is chosen, and the appropriate measure \eqref{inner} is introduced that makes the Hamiltonian constraint Hermitian. Now, if we want to work with the representation of the momentum operator like the one inspired by the scattering theory in the radial coordinates \cite{liboff1987introductory}, $\hat{P}=-i R^{-1}\partial_RR$, we can use the freedom offered by the operator ordering parameters to choose the measure as $R^2$ that makes the extension of $\hat{P}$ Hermitian on the physical states (normalized states),
	\begin{align}
	\braket{\psi|\hat{P}|\chi}&=\int_{0}^{\infty}dR\;R^2\psi^*\hat{P}\chi=-i\int_{0}^{\infty}dR\;\psi^*R\frac{\partial(R\chi)}{\partial R}\nonumber\\
	&=-i[R^2\psi^*\chi]^\infty_0+i\int_{0}^{\infty}dR\;R\frac{\partial(R\psi^*)} {\partial R}\chi=\braket{\hat{P}\psi|\chi},\text{ provided }R\psi(R,\tau)\rightarrow 0.
	\end{align}
	Thus the measure $R^2$ can accommodate a Hermitian Hamiltonian as well as a Hermitian momentum  on the space of physical states. In fact, it hold true for the case of the general measure $R^{1-a-2b}$ as well. Hence, the eigenstates of the momentum operator can be used to obtain the orthonormal incoming/outgoing modes. In this work, we will study the radiation profile of the dust cloud viz-a-viz incoming/outgoing modes defined through this approach.

	\subsection{Hermitian extension of the momentum operator in \texorpdfstring{$R^2$}{R2} measure space}\label{Rsq}

	The absence of self-adjoint extension for momentum operator on real half-line $\mathbb{R}^+$ is well documented in the literature see, e.g., Ref. \cite{Gieres_2000,Bonneau_2001,Gitman2012}. This suggests that the momentum operator on $\mathbb{R}^+$ is not a well defined observable in the quantum theory. In this subsection, we will argue that even though the momentum operator is not self-adjoint, we can still work with its Hermitian representation and its eigenfunctions.

	The eigenfunctions of a self-adjoint operator with different eigenvalues are orthogonal and yield a complete system of generalized vectors of the Hilbert space. This result might not hold for the operators, which are Hermitian but not self-adjoint \cite{Gieres_2000} and one has to show the eigenstates of such an operator are orthogonal explicitly.
	
	We will start with the discussion on the Hermiticity of the momentum operator and later check whether the eigenfunctions of the momentum operator form an orthogonal set of states or not. The boundary conditions that needs to be satisfied by the states for the Hermiticity of the Hamiltonian and momentum operator are,

    \begin{align}
		\braket{\psi|\hat{H}|\chi}-\braket{\hat{H}\psi|\chi}&=\biggr[R^{-a-2b}\bigg(\psi^*\frac{\partial\chi}{\partial R}-\frac{\partial\psi^*}{\partial R}\chi\bigg)\biggr]^\infty_0= 0,\label{BCh}\\
		\braket{\psi|\hat{P}|\chi}-\braket{\hat{P}\psi|\chi}&=\biggr[R^{1-a-2b}\psi^*\chi\biggr]^\infty_0= 0.\label{BC}
	\end{align}
	Let us have a function which has an asymptotic behavior $\psi(R)\rightarrow R^q$ as $R\rightarrow 0$ and $\psi(R)\rightarrow R^{q'}$ as $R\rightarrow\infty$. We will first write the constraint on the parameters $q$ and $q'$ arising from the demand that these functions are square integrable with measure $R^{1-a-2b}$, and satisfies the boundary conditions arising from the Hermiticity of the Hamiltonian and the momentum operator. We have summarized the boundary requirements for normalizability and Hermiticity of operators $\hat{H}$ and $\hat{P}$ in Table \ref{table}.

\begin{table}[H]
	\centering
	\bgroup
	\def\arraystretch{2.1}
	\begin{tabular}{ |c|c|c|}
		\hline
		\multirow{2}{*}{} & Behavior at boundaries & Condition for vanishing\\
		\hline
		\multirow{2}{*}{\textbf{Normalization}} & ${\displaystyle \lim_{R\rightarrow 0}R^{1-a-2b}\psi^*\psi\sim R^{2q+1-a-2b}}$ & $\underbrace{2q>a+2b-2}_{\text{For integral to vanish}}$\\\cline{2-3}
		&${\displaystyle \lim_{R\rightarrow\infty} R^{1-a-2b}\psi^*\psi\sim R^{2q'+1-a-2b}}$ &  $\underbrace{2q'<a+2b-2}_{\text{For integral to vanish}}$ \\
		\hline
		\multirow{2}{*}{\textbf{Hermitian $\hat{H}$}} & ${\displaystyle \lim_{R\rightarrow 0}R^{-a-2b}\bigg(\psi^*\frac{\partial\chi}{\partial R}-\frac{\partial\psi^*}{\partial R}\chi\bigg)}$  $\sim{\displaystyle \lim_{R\rightarrow 0}R^{2q-a-2b-1}}\quad$ & $\quad 2q>a+2b+1$ \\\cline{2-3}
		& ${\displaystyle \lim_{R\rightarrow \infty}R^{-a-2b}\bigg(\psi^*\frac{\partial\chi}{\partial R}-\frac{\partial\psi^*}{\partial R}\chi\bigg)}$  $\sim {\displaystyle \lim_{R\rightarrow\infty}R^{2q'-a-2b-1}}\quad$ & $\quad2q'<a+2b+1$ \\
		\hline
		\multirow{2}{*}{\textbf{Hermitian $\hat{P}$}} & ${\displaystyle \lim_{R\rightarrow 0} R^{1-a-2b}\psi^*\chi}$ $\sim{\displaystyle \lim_{R\rightarrow 0}R^{2q-a-2b+1}}\quad$ & $\quad2q>a+2b-1$ \\ 
		\cline{2-3}
		& ${\displaystyle \lim_{R\rightarrow\infty} R^{1-a-2b}\psi^*\chi}$ $\sim{\displaystyle \lim_{R\rightarrow\infty} R^{2q'-a-2b+1}}\quad$ & $\quad2q'<a+2b-1$ \\
		\hline
	\end{tabular}
	\egroup
	\caption{Conditions for normalization and Hermiticity of $\hat{H}$ and $\hat{P}$.}\label{table}
\end{table}
	
    For the case when the measure is $R^{2}$, the constraints on the parameters follow: (1) the normalization of the states are $q>-3/2$ and $q'<-3/2$, (2) the Hermiticity of the Hamiltonian constraint are $q>0$ and $q'<0$ and (3) the Hermiticity of the momentum operator are $q>-1$ and $q'<-1$. We are interested in the square-integrable states on which both momentum and Hamiltonian operator are Hermitian, which is achieved by demanding $q>0$ and $q'<-3/2$ for the $R^2$ measure and $q>(a+2b+1)/2$ and $q'<(a+2b-2)/2$ for the $R^{1-a-2b}$ measure. This choice ensures that the states are square integrable and the relevant boundary conditions are satisfied. This indicates that the Hamiltonian and momentum operators are Hermitian, and the expectation values of the operators computed for these states will be real. Therefore, we will work with the Hermitian extension of momentum operator and its eigenfunctions\footnote{When the operators in the quantum theory have the continuous spectrum, Rigged Hilbert space is the natural mathematical setting \cite{Madrid_2005}. Rigged Hilbert space, also called Gelfand triplet, is a triad of spaces $\Phi\subset\mathcal{H}\subset\Phi^\times$. Here $\Phi$ is the space of test functions or the Schwartz space $\mathcal{S}(\mathbb{R}^+)$ in this case, and $\Phi^\times$ is the antidual of $\Phi$ or the space of distributions. The generalized eigenstates of the operators belong to $\Phi^\times$ and its dual space $\Phi'$.}.
    
    The set of wave packets constructed for this model satisfies the boundary conditions (\ref{BCh},~\ref{BC}) that ensures the Hermiticity of the momentum operator. Now we have a Hermitian momentum operator which is not self-adjoint.  In Appendix \ref{or} we explicitly demonstrate that the eigenfunctions of the momentum operator are orthogonal with the measure chosen. Therefore the eigenstates with different eigenvalues are linearly independent, and they serve as valid incoming/outgoing modes. Hence, the eigenstates of the momentum operator can be used to obtain the orthonormal incoming/outgoing modes and we will study the radiation profile of the dust cloud with their help. We will first study the general characteristics for the measure $R^2$ and then generalize it for the measure $R^{1-a-2b}$.

	Since the eigenspace of $\hat{R}$ runs from $0$ to $\infty$, to obtain a representation of $\hat{P}$ which is Hermitian on the half-line $\mathbb{R}^+$, the momentum operator that is Hermitian with respect to the measure $R^2$ is,
	\begin{align}
	\hat{P}=-i R^{-1}\frac{\partial}{\partial R}R=-i\bigg( \frac{\partial}{\partial R}+\frac{1}{R}\bigg).\label{p}
	\end{align}
	
	The eigenfunctions of $\hat{P}$ are given by, $u_k(R)=e^{i kR}/R$, where $k\in\mathbb{R}$. For $R^2$ measure, we have to impose constraint $1-a-2b=2$ in \eqref{inner} or $a+2b=-1$ on operator ordering parameters. Using this constraint, we can eliminate one parameter. 

	For the Hermitian extension of the momentum operator on $\mathbb{R}^+$, we will see below that eigenfunctions with positive eigenvalue relate to incoming modes and eigenfunctions with negative eigenvalue relate to outgoing modes. We are interested in the projection of the wave packet along these orthogonal states. With this choice of measure, the Wheeler-DeWitt equation becomes,
	\begin{align}
	i\frac{\partial \Psi(R,\tau)}{\partial\tau}&=\hat{H}\Psi(R,\tau),\\
	\hat{H}=-\frac{1}{2}\hat{R}^{-b-1}\hat{P}\hat{R}^{2b+1}\hat{P}&\hat{R}^{-b-1}=\frac{1}{2}R^{-2-b}\frac{d}{dR}R^{2b+1}\frac{d}{dR}R^{-b}.
	\end{align}
	The stationary states of the Hamiltonian $\hat{H}\phi_E=-E\phi_E$ takes the form,
	\begin{align}
	\phi_E^1(R)=J_{\frac{2}{3}|b|}\left(\frac{2}{3}\sqrt{2E}R^{\frac{3}{2}}\right)\quad&,\quad	\phi_E^2(R)=Y_{\frac{2}{3}|b|}\left(\frac{2}{3}\sqrt{2E}R^{\frac{3}{2}}\right)\\
	\phi_{-E}^1(R)=I_{\frac{2}{3}|b|}\left(\frac{2}{3}\sqrt{2E}R^{\frac{3}{2}}\right)\quad&,\quad	\phi_{-E}^2(R)=K_{\frac{2}{3}|b|}\left(\frac{2}{3}\sqrt{2E}R^{\frac{3}{2}}\right)\\
	\phi_0^1=R^b\qquad&,\qquad\phi_0^2=R^{-b}.
	\end{align}
	Again using positive energy states, we can construct the wave packet using Poisson-like distribution
	\begin{align} 
	\psi(R,\tau)=\sqrt{3}\frac{R^{|b|}}{\sqrt{\Gamma(\frac{2}{3}|b|+1)}}\left(\frac{\frac{\sqrt{2\lambda}}{3}}{\frac{\lambda}{2}-i\tau}\right)^{\frac{2}{3}|b|+1}e^{-\frac{2R^3}{9\left(\frac{\lambda}{2}-i\tau\right)}}.
	\end{align}
	For the model in question, a general wave packet can be written in the form 
	\begin{align}
	\psi(R,\tau)=\int_{0}^\infty dE\;A(E)\phi_{E}(R)e^{i\tau E}.
	\end{align}
	The stationary states can be written as a linear combination of the eigenstates of the momentum operator,
	\begin{align}
	\phi_{E}(R)=\int_{-\infty}^{\infty}dk\;f(k,E)\frac{e^{i k R}}{R},
	\end{align}
	such that, the wave packet can be expressed as
	\begin{align}
	\psi(R,\tau)&=\int_{0}^\infty dE\int_{-\infty}^{\infty}dk\;A(E)\;f(k,E) \frac{e^{i k R}}{R}e^{i\tau E}\nonumber\\
	&=\int_{0}^\infty dE\int_{-\infty}^{\infty}dk\;\mathcal{A}(k,E) \frac{e^{i k R}}{R}e^{i\tau E},\text{ where }\mathcal{A}(k,E)=A(E)f(k,E)\nonumber\\
	&=\int_{0}^\infty dE\int_{0}^\infty dk\left(\mathcal{A}(k,E) \frac{e^{i k R+i\tau E}}{R}+\mathcal{A}(-k,E) \frac{e^{-i k R+i\tau E}}{R}\right).
	\end{align}	
	Looking at the phase factor, we see the incoming modes are identified with the momentum eigenfunctions with positive eigenvalue $u_{k,E}(\tau)\equiv e^{i k R+i\tau E}/R$ and outgoing modes with the negative eigenvalue $u_{-k,E}(\tau)\equiv e^{-i k R+i\tau E}/R$.
	The wave packet is decomposed in terms of incoming and outgoing modes. 
	\section{Infrared characteristics} \label{Infrared}
	Once we decompose the wave packet into the projection along the incoming $u_{k,E}(\tau)$ and outgoing $u_{-k,E}(\tau)$ modes, the projection of the wave packet along them will give the contribution of incoming/outgoing radiation through 
	\begin{align}
	\tilde{\psi}_k(\tau)&=\braket{u_{k,E}(\tau)|\psi(R,\tau)}=\int\;dR\;R^2\;\psi(R,\tau)\frac{e^{-i kR-i E\tau}}{R}\nonumber\\
	&=\frac{\sqrt{3}e^{-iE\tau} }{\sqrt{\Gamma(\frac{2}{3}|b|+1)}}\left(\frac{\frac{\sqrt{2\lambda}}{3}}{\frac{\lambda}{2}-i\tau}\right)^{\frac{2}{3}|b|+1}\int\;dR\;e^{-i\;kR}R^{1+|b|}e^{-\frac{2R^3}{9(\frac{\lambda}{2}-i\tau)}}\nonumber\\
	&=C(\tau)\int\;dR\;e^{-i\;kR}R^{1+|b|}e^{-\frac{2R^3}{9(\frac{\lambda}{2}-i\tau)}},\quad\text{with } C(\tau)=\frac{\sqrt{3}e^{-iE\tau}}{\sqrt{\Gamma(\frac{2}{3}|b|+1)}}\left(\frac{\frac{\sqrt{2\lambda}}{3}}{\frac{\lambda}{2}-i\tau}\right)^{\frac{2}{3}|b|+1}.\label{A1}
	\end{align}
	The wave packet in the $k$ space is normalized
	\begin{align}
	\int_{-\infty}^{\infty}dk\;\tilde{\psi}^*_k\tilde{\psi}_k=\int_{-\infty}^{\infty}dk\int\;dR\;R\;\psi^*(R,\tau)e^{i kR}\int\;dS\;S\;\psi(S,\tau)e^{-i kS}=2\pi.
	\end{align}
	Owing to the normalization, the mode functions decay when $k\rightarrow\infty$ and $|\tilde{\psi}_k(\tau)|^2$ gives the contribution of modes with wave number $k$ to the radiation profile at time $\tau$.
	
\begin{figure}[H]
		\centering
		\begin{tabular}{l c r}
			\includegraphics[scale=0.6]{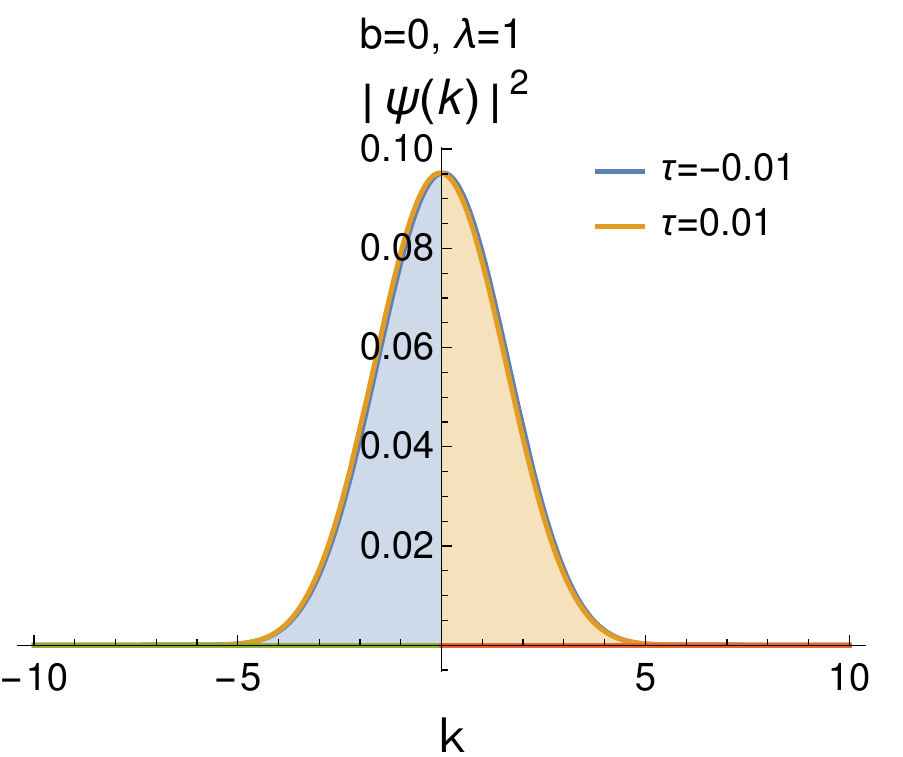} & \includegraphics[scale=0.6]{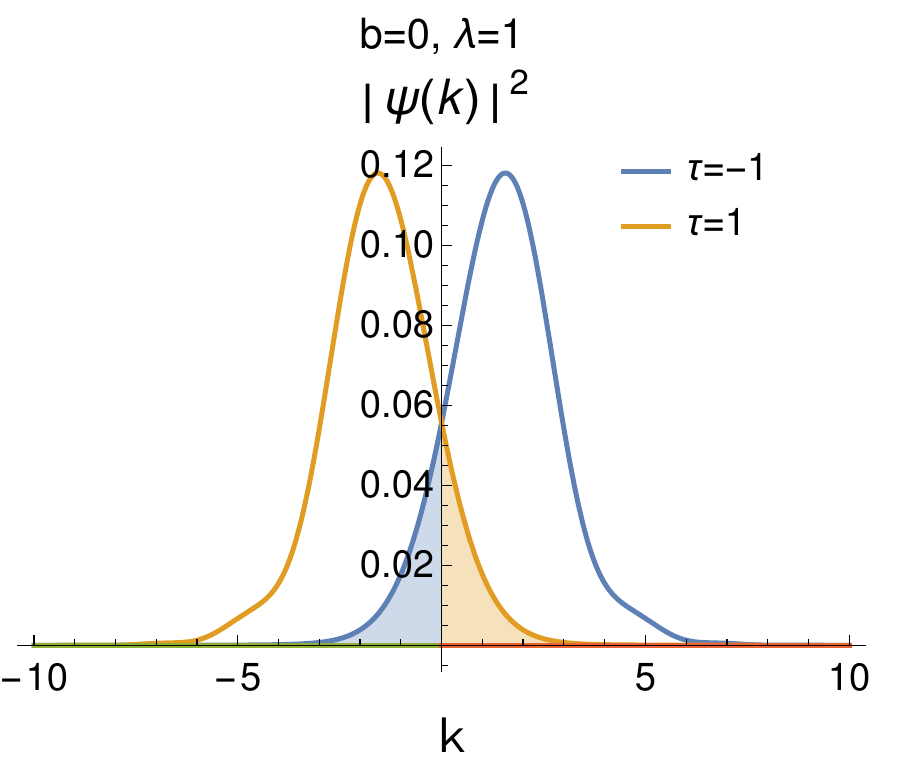} &  \includegraphics[scale=0.6]{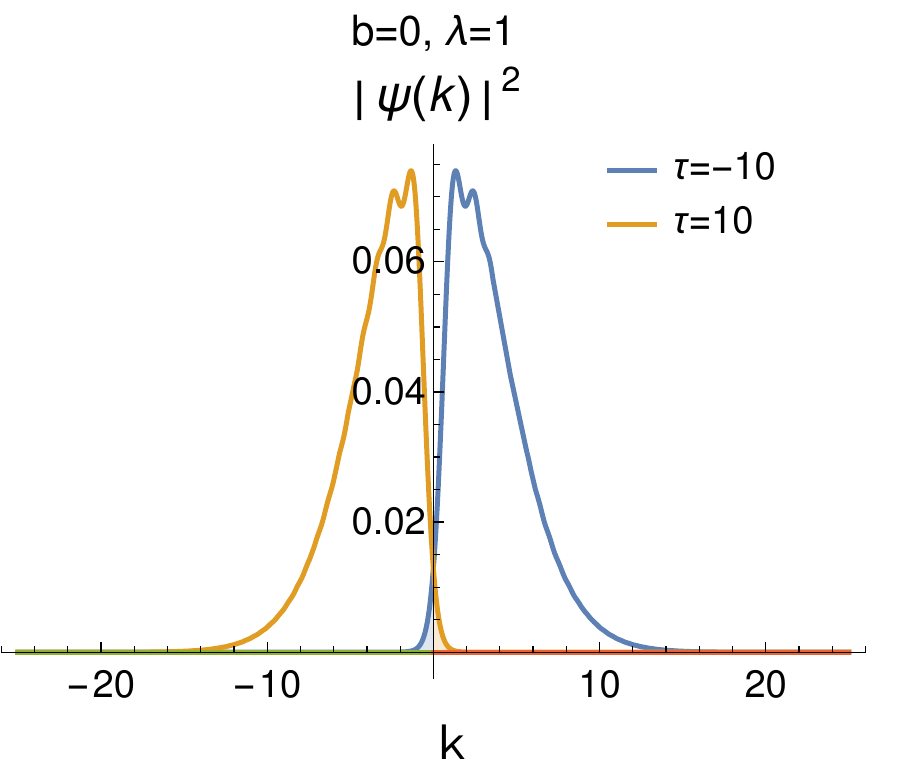}\vspace{0.5cm}\\
			\includegraphics[scale=0.6]{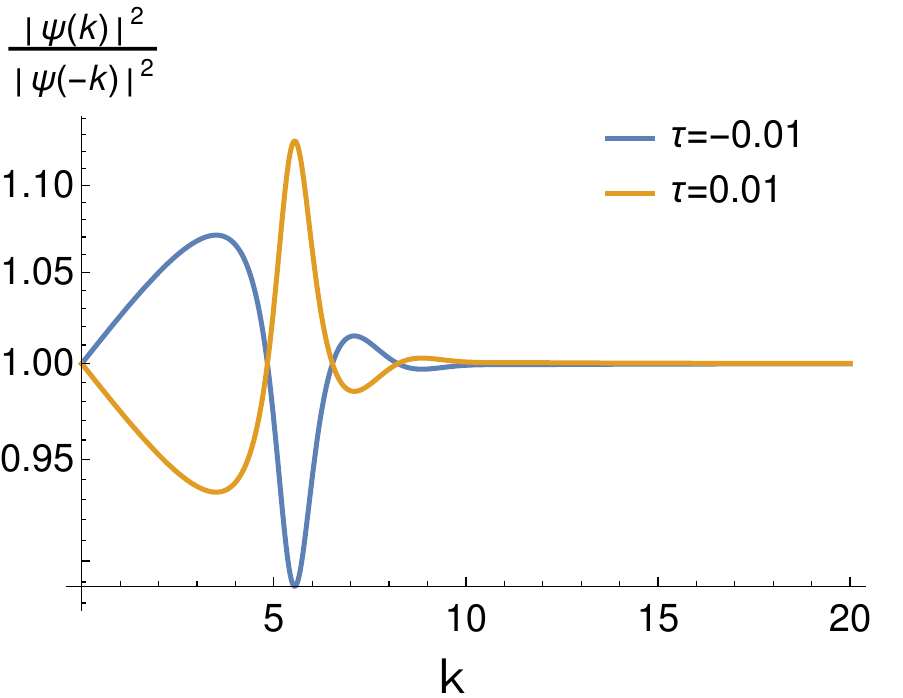} & \includegraphics[scale=0.6]{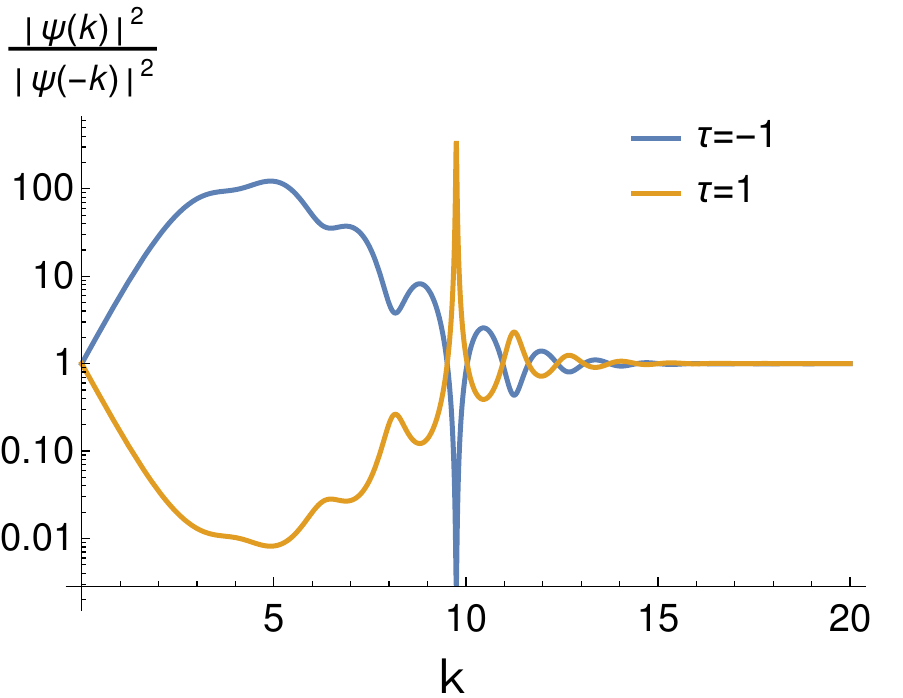} &  \includegraphics[scale=0.6]{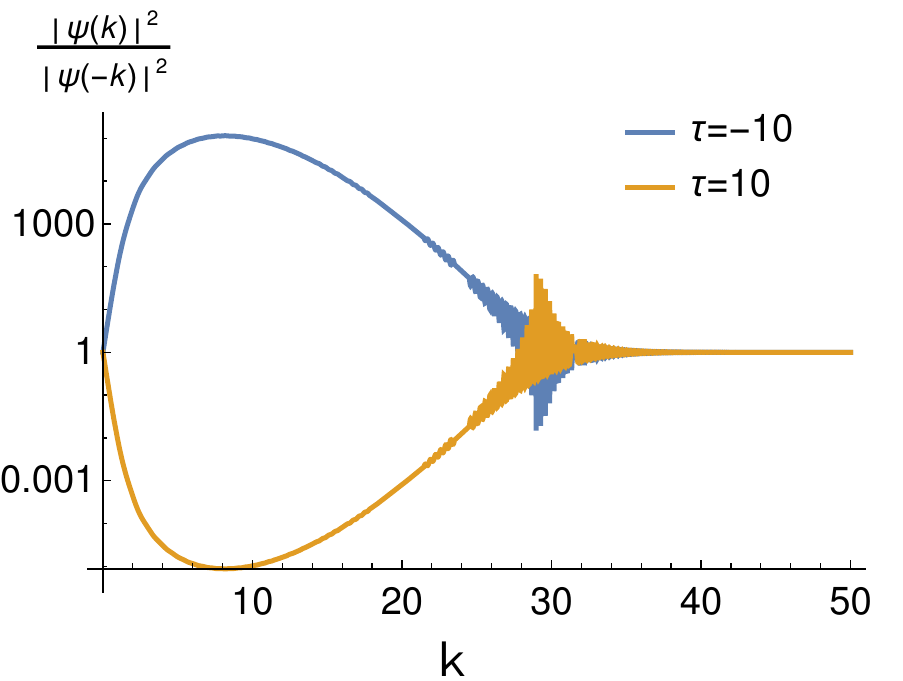}
		\end{tabular}
		\caption{Behavior of $|\tilde{\psi}_k(\tau)|^2$ and the ratio of incoming to outgoing modes at different times, with parameters specifying narrow wave packet. The shaded region gives the contribution of outgoing radiation in collapsing phase (light blue) and incoming radiation in an expanding phase (light orange).}
		\label{plot1}
	\end{figure}

	At $\tau=0$, Eq. \eqref{A1} can be written as
	\begin{align}
	\tilde{\psi}_k(0)=C(0)\left(\int\;dR\;\cos\left(kR\right)R^{1+|b|}e^{-\frac{4R^3}{9\lambda}}-i \int\;dR\;\sin\left(kR\right)R^{1+|b|}e^{-\frac{4R^3}{9\lambda}}\right),
	\end{align}
	clearly $|\tilde{\psi}_k(0)|^2=|\tilde{\psi}_{-k}(0)|^2$. Thus at the point of classical singularity, the number of incoming modes becomes equal to the number of outgoing modes for all $k$. On the other hand, for finite $\tau$, Eq. \eqref{A1} can be cast into the form,
	\begin{align}
	|\tilde{\psi}_k(\tau)|^2&=|C(\tau)|^2\int dR\int dS\;e^{-i k(R-S)}(RS)^{1+|b|}e^{-\frac{2R^3}{9\left(\frac{\lambda}{2}-i\tau\right)}-\frac{2S^3}{9\left(\frac{\lambda}{2}+i\tau\right)}}.
	\end{align}
	Here we can see, taking $\tau\rightarrow-\tau$ is equivalent to taking $k\rightarrow-k$. Thus the ratio of incoming to outgoing modes $r_k(\tau)=|\tilde{\psi}_k(\tau)|^2/|\tilde{\psi}_{-k}(\tau)|^2$ flips when the bounce happens i.e. $r_k(\tau)=[r_k(-\tau)]^{-1}$, see Fig. \ref{plot1}. The ratio greater than one implies incoming modes are dominating in that regime. The difference between the number of incoming and outgoing modes at any instant of time and at a fix $k$ can be written as,
	\begin{align}
	\delta_k(\tau)&=|\tilde{\psi}_k(\tau)|^2-|\tilde{\psi}_{-k}(\tau)|^2=|C(\tau)|^2\int dR\int dS\;(e^{-i k(R-S)}-e^{i k(R-S)})(RS)^{1+|b|}e^{-\frac{2R^3}{9\left(\frac{\lambda}{2}-i\tau\right)}-\frac{2S^3}{9\left(\frac{\lambda}{2}+i\tau\right)}}\nonumber\\
	&=-2i|C(\tau)|^2\int dR\int dS\;\sin(k(R-S))(RS)^{1+|b|}e^{-\frac{2R^3}{9\left(\frac{\lambda}{2}-i\tau\right)}-\frac{2S^3}{9\left(\frac{\lambda}{2}+i\tau\right)}}.
	\end{align}
	Because $\underset{k\rightarrow\infty}{\lim}\sin(kx)=x\delta(x)=0$, the  $\delta_k(\tau)$ vanishes when $k\rightarrow\infty$. Thus we expect the ratio to approach unity for large $k$, see Fig. \ref{plot1}. On the other hand, for $k\rightarrow0$ we have
	\begin{align}
	\delta_k(\tau)&=-2i\;k|C(\tau)|^2\int dR\int dS\;(R-S)(RS)^{1+|b|}e^{-\frac{2R^3}{9\left(\frac{\lambda}{2}-i\tau\right)}-\frac{2S^3}{9\left(\frac{\lambda}{2}+i\tau\right)}}\nonumber\\
	&=-2i\;k|C(\tau)|^2\left(I_{2+b}(\tau,\lambda)\;I_{1+b}^*(\tau,\lambda)-I_{2+b}^*(\tau,\lambda)\;I_{1+b}(\tau,\lambda)\right),
	\end{align}
	where,
	\begin{align}
	I_n(\tau,\lambda)=&\int dR\;R^n e^{-\frac{2R^3}{9\left(\frac{\lambda}{2}-i\tau\right)}}=\frac{1}{3}\Gamma\left[\frac{n+1}{3}\right]\left(\frac{2}{9\left(\frac{\lambda}{2}-i\tau\right)}\right)^{-\frac{n+1}{3}},\\
	\left(I_{2+b}\;I_{1+b}^*-I_{2+b}^*\;I_{1+b}\right)=&2i\left(\frac{1}{9}\right)^{\frac{|b|+8}{3}}\Gamma\left[\frac{|b|+3}{3}\right]\Gamma\left[\frac{|b|+2}{3}\right]\left(\frac{\lambda^2}{4}+\tau^2\right)^{\frac{2|b|+5}{6}}\sin\left(\frac{1}{3}\tan^{-1}\left(\frac{-2\tau}{\lambda}\right)\right)
	\end{align}
	The difference between the outgoing and the incoming modes for $k\rightarrow0$  reads
	\begin{align}
	\delta_k(\tau)=4k\;|C(\tau)|^2\left(\frac{1}{9}\right)^{\frac{|b|+8}{3}}&\Gamma\left[\frac{|b|}{3}+1\right]\Gamma\left[\frac{|b|+2}{3}\right]\left(\frac{\lambda^2}{4}+\tau^2\right)^{\frac{2|b|+5}{6}}\sin\left(\frac{1}{3}\tan^{-1}\left(\frac{-2\tau}{\lambda}\right)\right).
	\end{align}
	Since $C^*(\tau)=C(-\tau)$, we can see the function $\delta_k$ is an odd function of $\tau$ and for $\tau>0$, $\delta_k(\tau)<0$. Therefore, the incoming modes dominate in the contracting branch, and after the bounce, outgoing modes dominate in the expanding branch for small wave numbers.
	
	We choose a parameter set that represents a narrow wave packet localized along the classical trajectory of the moderate energy dust shell. These plots in Fig. \ref{plot1} describe the contribution of different wave number modes to the dust shell at different time slices, which represents very early in the collapse ($\tau<<0$), very late in the expanding phase ($\tau>>0$), and near the classical singularity ($\tau\sim0$).

	Early on during the collapse, the majority contribution comes from incoming modes along with a tiny fraction of outgoing modes of small wave number contributing as well, shown by the shaded region in Fig. \ref{plot1}. As we approach $\tau=0$, the contribution coming from outgoing modes in the collapsing branch keeps on rising, and at the classical singularity, the number of incoming modes becomes equal to the number of outgoing modes. At the start of the expansion, the fraction of incoming modes is sizable, and it keeps on decreasing as the shell expands. At the later stage of expansion, the dust shell is mainly comprised of outgoing modes with a very small contribution of incoming modes at small wave number or large wavelength. Also, most of the contribution to incoming/outgoing modes in the expanding/collapsing phase comes from a relatively small wave number regime.

	The ratio of incoming to outgoing modes for the collapsing branch starts from unity at $k=0$ and starts increasing, acquiring a maximum, then decreases and oscillates before settling again at unity for some finite wave number. As we approach the classical singularity, the magnitude of maximum keeps on decreasing. At $\tau=0$, this ratio is unity for all values of $k$. This behavior inverts when we go from the collapsing to the expanding branch. Again, the ratio starts from one and decreases, attains a minimum, and then increases and oscillates before settling at unity again. In the collapsing branch and close to the singularity i.e., $\tau\sim0$, there is a crossover in the plot when the contribution from the outgoing radiation exceeds the contribution coming from the incoming radiation for a window of wave numbers. A reflected behavior is observed in the expanding branch as well. 

	If an observer analyzes the small wave number $k$, i.e. infrared regime of the dust shell, there will be an instantaneous flip from largely incoming radiation to largely outgoing radiation as we go from the collapsing to the expanding branch. An observer looking at the dust cloud at small wave number or the large wavelength part of the radiation will know if the bounce has happened instantaneously, much before any information comes about, provided that the amplitudes are observable in the infrared sector, which we will estimate in the next subsection.

\begin{figure}[H]
	\centering
	\begin{tabular}{c c}
		\includegraphics[scale=0.62]{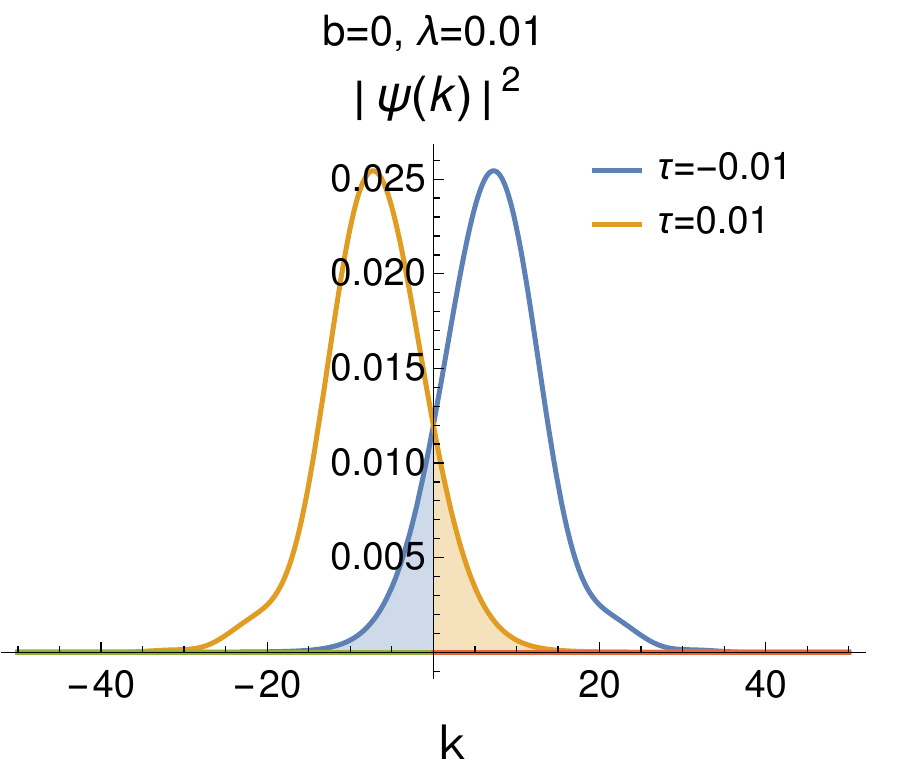}\hspace{1.5cm} & \includegraphics[scale=0.62]{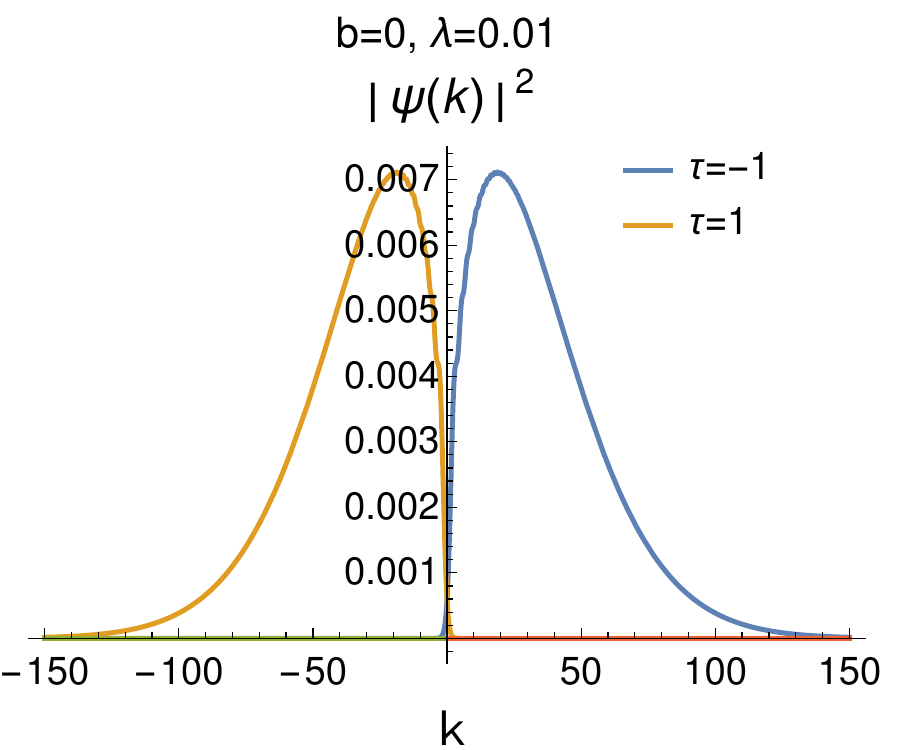} \\ \includegraphics[scale=0.62]{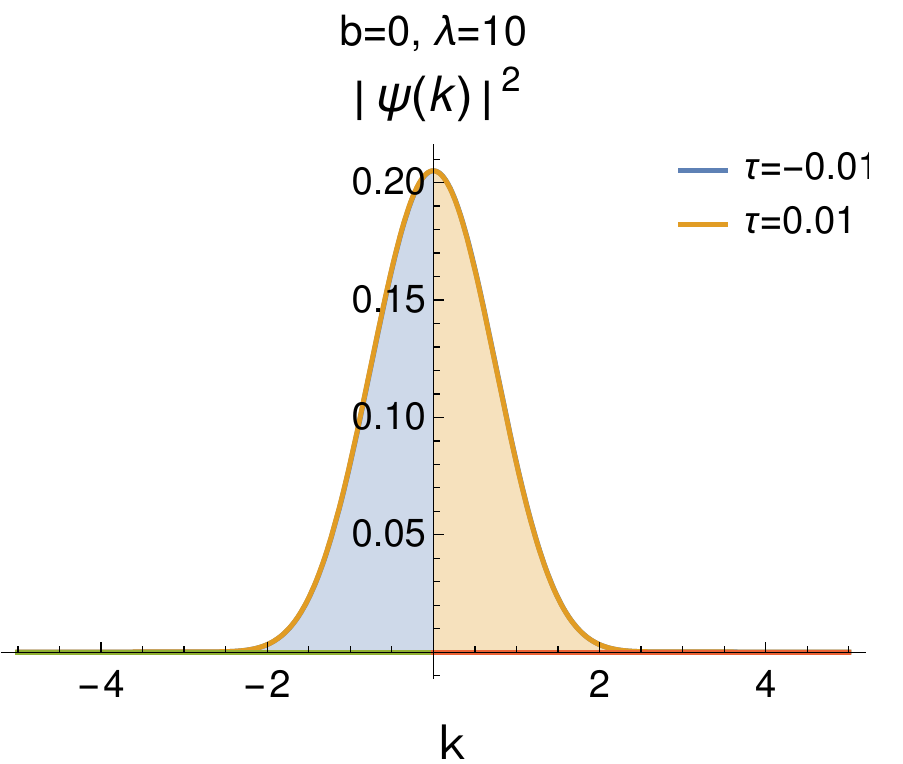} \hspace{1.5cm} & \includegraphics[scale=0.62]{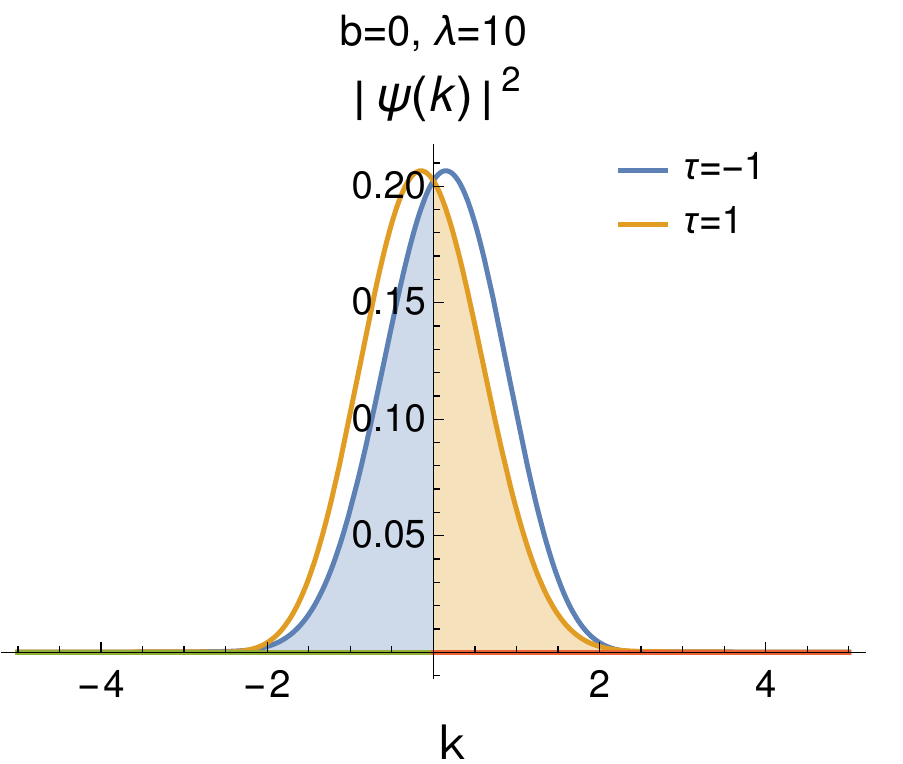} 
	\end{tabular}
	\caption{Behavior of $|\tilde{\psi}_k(\tau)|^2$ and ratio of incoming to outgoing modes at different times, with parameters specifying a sharper ($\lambda=0.01$) wave packet representing a high energy dust shell and broader ($\lambda=10$) wave packet representing a low energy dust shell. The shaded region gives the contribution of outgoing radiation in the collapsing phase (light blue) and incoming radiation in the expanding phase (light orange).}
	\label{plot2}
\end{figure}

	The parameter set $b=0$ and $\lambda=1$ represents a sharp wave packet localized along the classical trajectory while the energy of the dust shell is inversely proportional to $\lambda$, see Eq. \eqref{energy}.  The sharper wave packet represents a shell with large energy, and the broader wave packet represents a dust shell with small energy. Let us analyze the behavior of $|\tilde{\psi}_k(\tau)|^2$ near the singularity for sharper wave packet e.g. $\lambda=0.01$ and broader wave packet, e.g. with $\lambda=10$.

	We can see for a sharper wave packet representing a dust shell with high energy, the contribution of incoming (outgoing) radiation in expanding (collapsing) phase is small, away from the singularity, e.g. at $\tau=\pm 1$ and the radiation profile mostly consists of outgoing (incoming) radiation in the expanding (collapsing) phase. On the other hand, for dust shells with low energy, the contribution of incoming/outgoing modes is comparable even away from the singularity, e.g. again at $\tau=\pm 1$. Moreover, there is a significant contribution of the infrared modes that persists even in the stages of dust shell away from the singularity. Thus the infrared regime of the low energy dust shell dynamics might be more suitable to study the quantum signatures. We now estimate the amplitude in the infrared sector.
	
	\subsection{Infrared estimate of the bounce radius}\label{infrared}
	
\begin{figure}[H]		
		\begin{tabular}{c c}
			\includegraphics[scale=0.60]{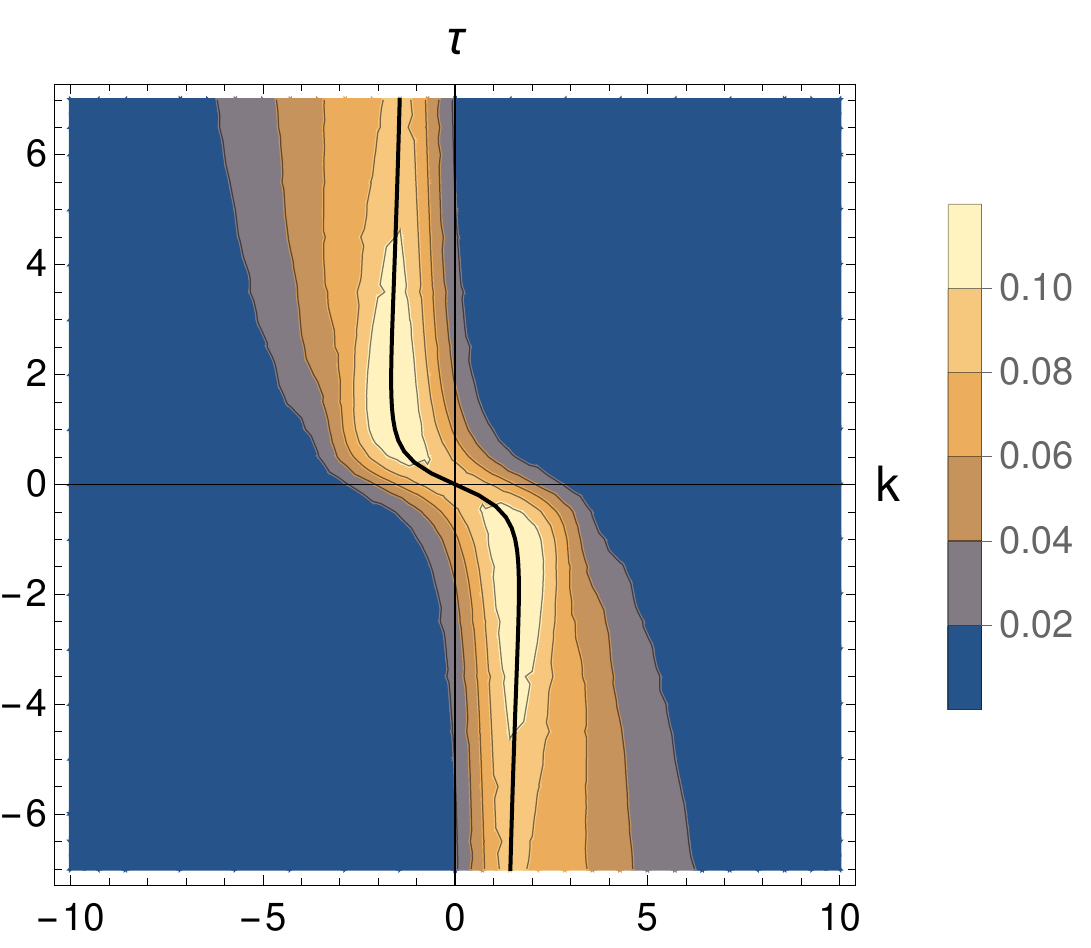} \hspace{1.5cm}& \includegraphics[scale=0.50]{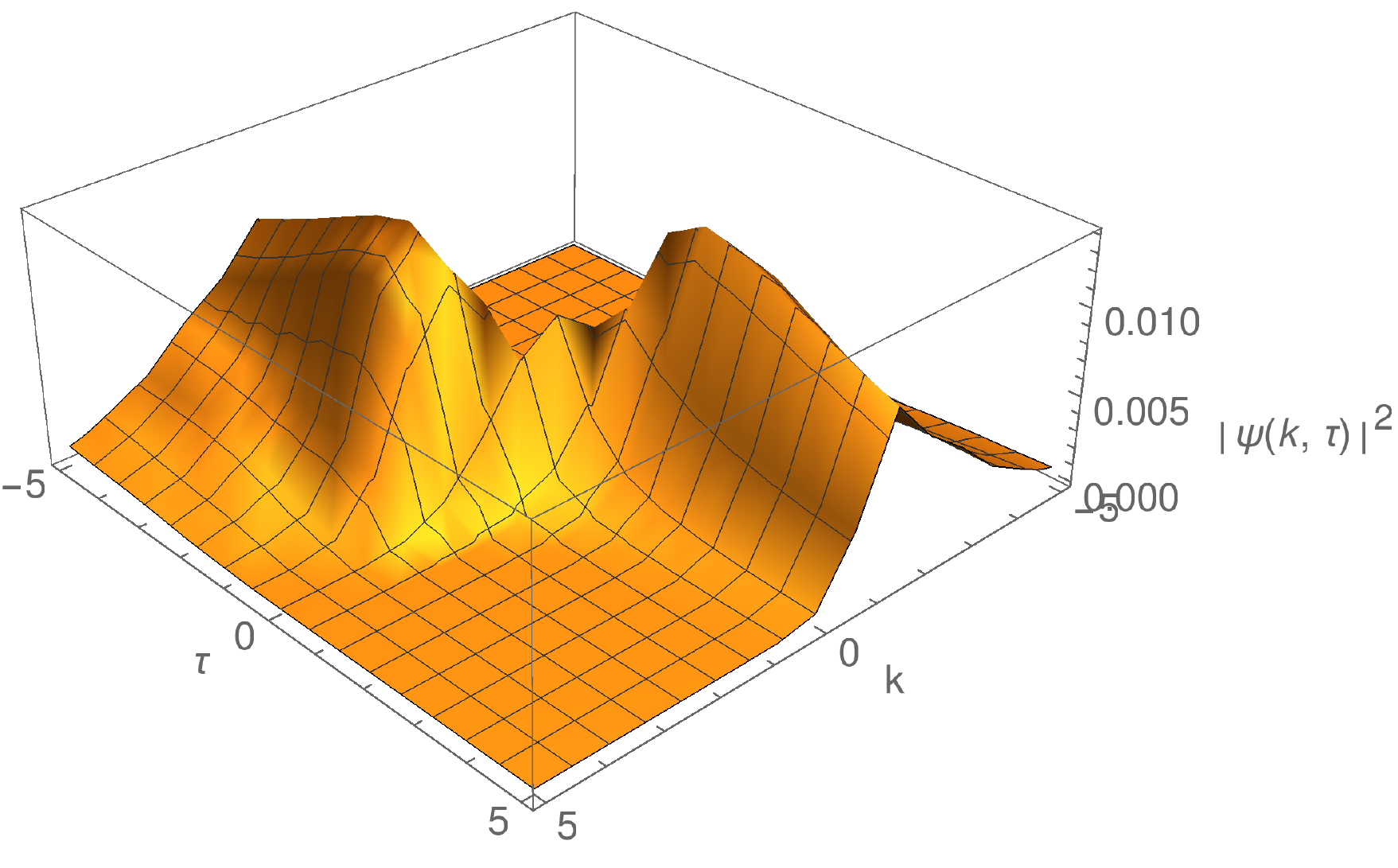}  
		\end{tabular}
		\caption{Plot of $|\psi_k(\tau)|^2$ for $b=0$ and $\lambda=1$. $k=0$ and $\tau=0$ is a local maximum of the function. Th solid black line in contour plot is the locus of points on which $|\psi_k(\tau)|^2$ is maximum at fixed $\tau$. Global maxima of this function are at $k=\pm 1.65163$ and $\tau=\mp 1.53312$.}\label{3dplot}
	\end{figure}

	The integral \eqref{A1} has an asymptotic expression for $k\rightarrow0$,
	\begin{align}
	\tilde{\psi}_0(\tau)=\frac{2^{\frac{|b|}{3}-\frac{1}{3}} \lambda^{\frac{1}{2}+\frac{|b|}{3}}  \Gamma \left(\frac{|b|+2}{3}\right)  (\lambda -2 i \tau )^{-\frac{|b|+1}{3}}}{\sqrt[6]{3} \pi  \sqrt{\Gamma \left(\frac{2|b|+1}{3}\right)}} +{\cal O}(k).
	\end{align}
	We are interested in the properties of absolute squared function
	\begin{align}
	f(\tau)=|\tilde{\psi}_0&(\tau)|^2=\frac{2^{\frac{2|b|-2}{3}} \lambda^{1+\frac{2|b|}{3}}  \Gamma \left(\frac{|b|+2}{3}\right)^2  (\lambda^2 +4 \tau^2 )^{-\frac{|b|+1}{3}}}{\sqrt[3]{3} \pi  \Gamma \left(\frac{2|b|+1}{3} \right)}=C(\lambda,b)(\lambda^2 + 4 \tau^2 )^{-\frac{|b|+1}{3}},\label{smallk}\\
	f'(\tau)=&\;C(\lambda,b)\left(-\frac{|b|+1}{3}\right)(\lambda^2 + 4 \tau^2 )^{-\frac{|b|+1}{3}-1}8\tau\implies f'(0)=0,\\
	f''(\tau)=&-\frac{8}{3}(|b|+1)C(\lambda,b)(\lambda^2 + 4 \tau^2 )^{-\frac{|b|+4}{3}}+\frac{64}{3}(b+1)\tau\left(\frac{|b|+1}{3}+1\right)C(\lambda,b)(\lambda^2 + 4 \tau^2 )^{-\frac{|b|+7}{3}},\nonumber\\
	f''(0)=&-\frac{8}{3}(|b|+1)C(\lambda,b)\lambda^{-\frac{ 2|b|+8}{3}}.
	\end{align}
	This function has a maximum at $\tau=0$ which means that the contribution of the small wave number modes peaks at the classical singularity. Therefore, in the infrared regime, the emission profile peaks at the bounce point. Moreover, for $\tau=0$, the integral \eqref{A1} is expanded up to second order in $k$ as
	\begin{align}
	\tilde{\psi}_k(0)=&\frac{2^{\frac{|b|-8}{3}} \sqrt[6]{\lambda } \left(-9 k^2 \lambda ^{2/3} \Gamma \left(\frac{|b|+4}{3}\right)-6\ 2^{2/3} \sqrt[3]{3} i k \sqrt[3]{\lambda } \Gamma \left(\frac{|b|}{3}+1\right)+4 \sqrt[3]{2} 3^{2/3} \Gamma \left(\frac{|b|+2}{3}\right)\right)}{3^{5/6} \sqrt{\pi } \sqrt{\Gamma \left(\frac{1}{3} (2|b|+1)\right)}},\\
	|\tilde{\psi}_k(0)|^2=&\frac{2^{\frac{2(|b|-8)}{3}} \!\!\left(27 k^4 \lambda ^{5/3} \Gamma \left(\frac{|b|+4}{3}\right)^2\!\!+24 \sqrt[3]{2} 3^{2/3} k^2 \lambda  \left(\Gamma \left(\frac{|b|+3}{3}\right)^2\!\!-\!\!\Gamma \left(\frac{|b|+2}{3}\right) \Gamma \left(\frac{|b|+4}{3}\right)\right)+16\ 2^{2/3} \sqrt[3]{3} \sqrt[3]{\lambda } \Gamma \left(\frac{|b|+2}{3}\right)^2\right)}{3^{2/3} \pi  \Gamma \left(\frac{2|b|}{3}+\frac{1}{3}\right)}\nonumber\\
	=&g(k).
	\end{align}
	We are interested in the properties of function $g(k)$ as $k\rightarrow 0$,
	\begin{align}
	\frac{dg(k)}{dk}&=\frac{2^{\frac{2 (|b|-8)}{3}} \left(108 k^3 \lambda ^{5/3} \Gamma \left(\frac{|b|+4}{3}\right)^2+48 \sqrt[3]{2} 3^{2/3} k \lambda  \left(\Gamma \left(\frac{|b|+3}{3}\right)^2-\Gamma \left(\frac{|b|+2}{3}\right) \Gamma \left(\frac{|b|+4}{3}\right)\right)\right)}{3^{2/3} \pi  \Gamma \left(\frac{2|b|}{3}+\frac{1}{3}\right)}\!\!\implies\!\! \frac{dg(k)}{dk}\biggr|_{k=0}\!\!\!\!\!=0,\\
	\frac{d^2g(k)}{d^2k}&=\frac{2^{\frac{2 (|b|-8)}{3}} \left(324 k^2 \lambda ^{5/3} \Gamma \left(\frac{|b|+4}{3}\right)^2+48 \sqrt[3]{2} 3^{2/3} \lambda  \left(\Gamma \left(\frac{|b|+3}{3}\right)^2-\Gamma \left(\frac{|b|+2}{3}\right) \Gamma \left(\frac{|b|+4}{3}\right)\right)\right)}{3^{2/3} \pi  \Gamma \left(\frac{2|b|}{3}+\frac{1}{3}\right)}\!\!\!\implies\!\!\!\frac{d^2g(k)}{d^2k}\biggr|_{k=0}\!\!\!\!\!<0,
	\end{align}
	suggesting this function acquires local maximum at $k=0$ and $\tau=0$ 
	as can be seen in Fig. \ref{3dplot}.
	
	Further, the value of $|\tilde{\psi}_k(\tau)|^2$ at the local maximum can be evaluated to be
	\begin{align}
	|\tilde{\psi}_0(0)|^2=\frac{2^{\frac{2 |b|-2}{3}} \Gamma \left(\frac{|b|+2}{3}\right)^2  }{\sqrt[3]{3} \pi  \Gamma \left(\frac{2 |b|+1}{3}\right)}\lambda^{\frac{1}{3}}.
	\end{align}
	Since the energy of the dust shell is inversely proportional to $\lambda$, it means that the fraction of incoming/outgoing modes at large wavelength is proportional to the energy of the dust shell via $\bar{E}^{-1/3}$. The low energy dust shell is suitable for studying the quantum signatures in the infrared regime of the radiation profile. From Eq. \eqref{A6}, we see that the expectation value of the areal radius at $\tau=0$ is also proportional to $\lambda^{1/3},$ leading to
	\begin{align}
	|\tilde{\psi}_0(0)|^2=\frac{2^{\frac{2 |b|-5}{3}}3^{\frac{2}{3}} \Gamma \left(\frac{|b|+2}{3}\right)^2 \Gamma(\frac{2|b|+4}{3})}{\sqrt{2} \pi  \Gamma \left(\frac{2 |b|+1}{3}\right)\Gamma(\frac{2}{3}|b|+1)} \bar{R}(0), \label{k0t0}
	\end{align}
	i.e., $|\psi_{k\rightarrow 0}(0)|^2\propto\bar{R}(0)$, which means that the number density of incoming/outgoing modes at small wave numbers is sensitive to the minimal size of dust shell. Thus the infrared regime of the radiation profile provides a direct estimation of the bounce radius. This object is also sensitive to the parameter $b$ and due to the constraint $a+2b+1=0$ effectively on $a$, but it is not possible to establish whether the dependency is a direct artifact of operator ordering ambiguity in the model or whether it is coming from the shape of energy distribution used to construct the wave packet.

	\section{Hermitian momentum in \texorpdfstring{\(R^{1-a-2b}\)}{R} measure space}\label{Rgen}
	In the previous subsection, we have used the representation of the momentum operator and the measure inspired by quantum scattering theory in spherical polar coordinates \cite{liboff1987introductory}. In this subsection, we wish to check if the features obtained in the previous subsection remain valid for the general measure and the operator ordering of momentum as well. The Hermitian extension of the momentum operator in $L^2(\mathbb{R}^+,R^{1-a-2b}dR)$ is given by,
	\begin{align}
	\hat{P}=-i R^{-\frac{1}{2}(1-a-2b)}\frac{\partial}{\partial R}R^{\frac{1}{2}(1-a-2b)}=-i\hbar\left[\frac{\partial}{\partial R}+\frac{1-a-2b}{2R}\right].\label{mo}
	\end{align}
	The eigenstates of momentum operator with eigenvalue $k$ are $u_k=e^{i kR}/R^{\frac{1}{2}(1-a-2b)}$. Orthogonality of these eigenstates is established in the appendix \ref{or}. As is discussed in the previous subsection, incoming modes are associated with the momentum eigenstates with positive eigenvalue $u_{k,E}(\tau)\equiv e^{ikR+i\tau E}/R^{\frac{1}{2}(1-a-2b)}$, and outgoing modes are associated with negative eigenvalue eigenstates $u_{k,E}(\tau)\equiv e^{-ikR+i\tau E}/R^{\frac{1}{2}(1-a-2b)}$. Again we are interested in the projection of the wave packet \eqref{wp} along these modes,
	\begin{align}
	\psi_k(\tau)&=\braket{u_{k,E}|\psi}=e^{-i E \tau} \int_{0}^{\infty}\frac{e^{-i kR}}{R^{\frac{1}{2}(1-a-2b)}}\psi(R,\tau)R^{1-a-2b}dR\nonumber\\
	&=\frac{\sqrt{3}e^{-i E \tau} }{\sqrt{\Gamma\left(\frac{1}{3}|1+a|+1\right)}}\left(\frac{\sqrt{2 \lambda }}{3 \left(\frac{\lambda }{2}-i \tau \right)}\right)^{\frac{1}{3}|a+1|+1}\int_{0}^{\infty}e^{-i kR}R^{\frac{1}{2}(2+|1+a|)}e^{-\frac{2 R^3}{9 \left(\frac{\lambda }{2}-i \tau \right)}}dR.
	\end{align}
	
	Here, we can see that the projection is independent of the operator ordering parameter $b$. We will show in Sec. \ref{Obsdep}, it is a free parameter of the theory as the expectation value of observables in the general wave packet are independent of the parameter $b$. The form of the integral is similar to what we have had in the previous case \eqref{A1}. Thus we can expect the behavior of the mode function to be reminiscent of what we have seen in the last section.\\
	The analytical discussion also follows along the same lines. At the classical singularity, the number of incoming modes is equal to the number of outgoing modes. For $k\rightarrow0$, incoming modes dominate in the collapsing branch, and outgoing modes dominate in the expanding branch. For this case as well, the ratio does not change if we take $k\rightarrow-k$ and $\tau$$\rightarrow-\tau$, i.e., the ratio of incoming to outgoing modes inverts as we go from collapsing branch to expanding branch.
	
	\begin{figure}[H]
		\centering
		\begin{tabular}{l c r}
			\includegraphics[scale=0.62]{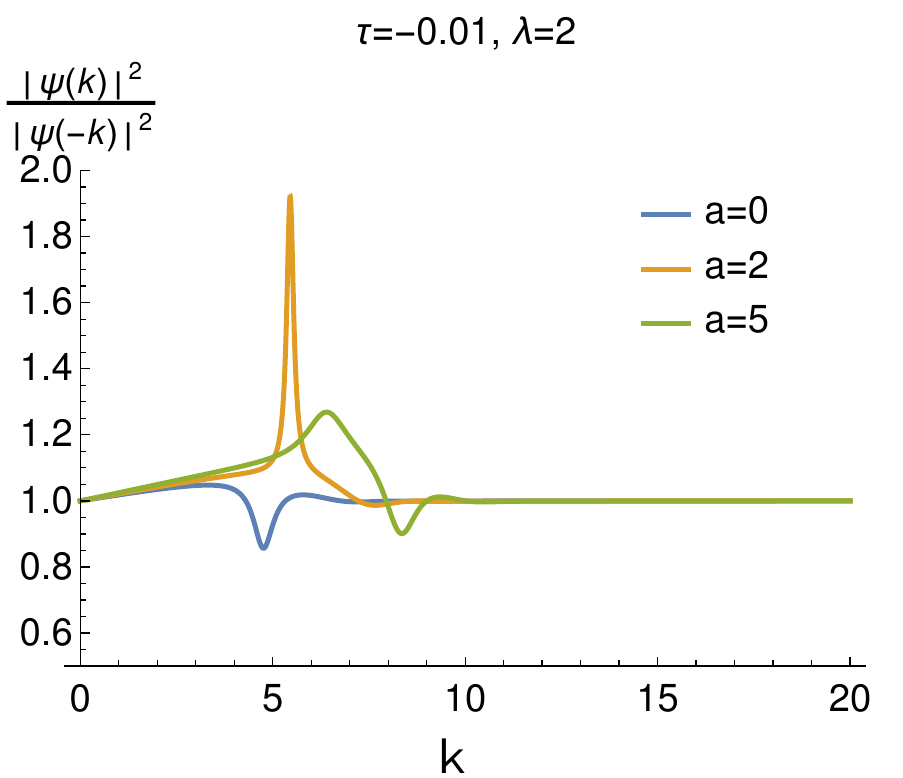} & \includegraphics[scale=0.62]{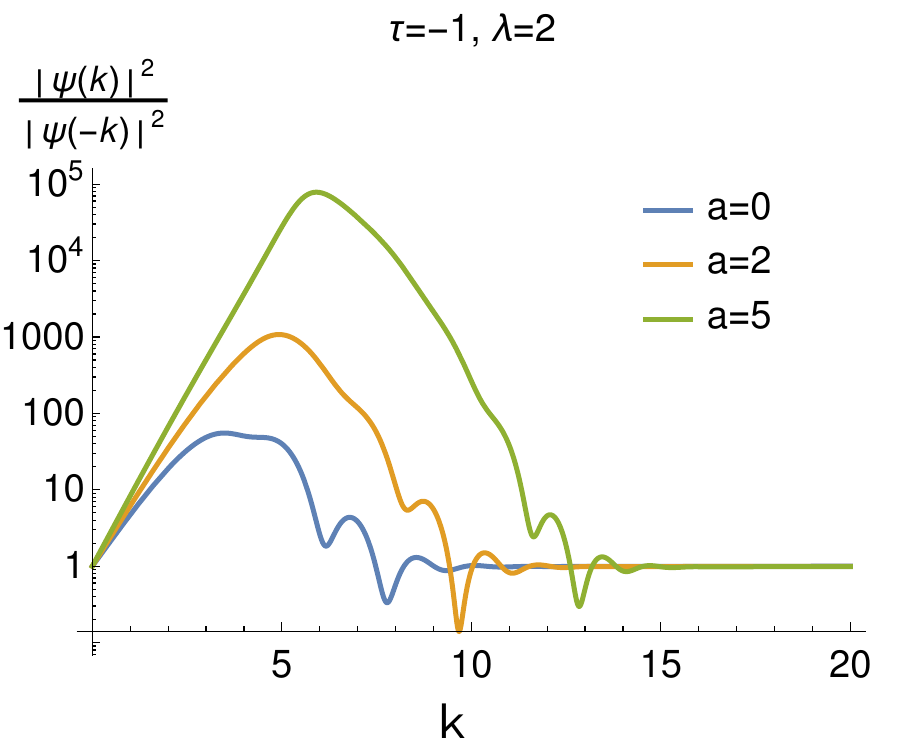} &  \includegraphics[scale=0.62]{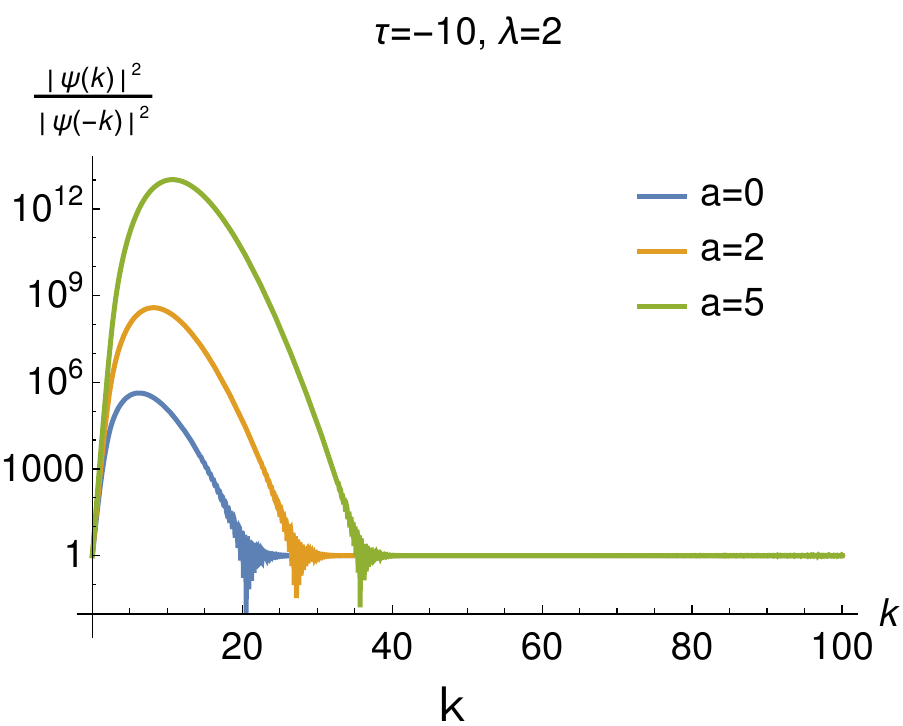}
		\end{tabular}
		\caption{Ratio of incoming to outgoing modes at various stages of collapsing phase with parameters specifying narrow localized wave packet and different operator ordering parameter $a$.}
		\label{plot4}
	\end{figure}

	The ratio of incoming to outgoing modes follows a similar trend as is observed in the previous subsection. Here as well, in the contracting branch, the ratio starts from unity and starts increasing to reach a maximum, decreases to an oscillatory regime and finally settle at unity for finite wave number. The same trend is observed in the expanding branch as well, although not shown in the plots. Moreover, the magnitude of the maximum keeps on decreasing as we keep on approaching the singularity. The same trend is observed for different values of parameter $a$ as well. Therefore these features appear for the general measure and the operator ordering of momentum operator as well, and the information of parameter $a$ gets transferred to the infrared regime as collapse progresses.
	\section{Observable's dependency on Operator Ordering Parameters}\label{Obsdep}
	As we have seen in Sec. \ref{infrared}, changing the operator ordering parameters brings about a change in $|\psi_{k}|^2$, but the prescription used to obtain wave packet \eqref{wp} restricts us to infer whether these changes are due to the change in the distribution or do they imply a genuine dependence on operator ordering parameters. Further, the space of initial conditions (configurations) can in principle be a function of $a$ and $b$ leaving representation of $\hat{P}$ fixed. To establish whether there is a concrete observable signature of operator ordering parameters, it is imperative to check whether the expectation value of observables in the general wave packet is a function of these parameters or not. We will work with a general wave packet constructed from positive energy stationary states,
	\begin{align}
	\Psi(R,\tau)=\int dE\;A(E)\;e^{i E\tau}R^{\frac{1}{2}(1+a+2b)} J_{\frac{1}{3}|1+a|}\left(\frac{2}{3}\sqrt{2E}R^{\frac{3}{2}}\right).
	\end{align}
	
	In further analysis, we will write $\frac{1}{3}|1+a|=\nu_a$ and $\frac{2}{3}\sqrt{2E}=\lambda_E$. We are interested in the expectation values of general phase space observables in this wave packet.
	
	\begin{enumerate}
		
		\item $\mathbf{\hat{R}^n:}$ We will first consider expectation value of arbitrary power of the areal radius operator,
		\begin{align}
		\braket{\Psi|\hat{R}^n|\Psi}&=\int_{0}^{\infty}dR\;R^{1-a-2b}\Psi^*\hat{R}^n\Psi,\nonumber\\
		&=\int\;dEdE'A^*(E)A(E')e^{-i(E-E')\tau}\int\;dR\;R^{n+1-a-2b}R^{1+a+2b} J_{\nu_a}\left(\lambda_{E'}R^{\frac{3}{2}}\right)J_{\nu_a}\left(\lambda_E R^{\frac{3}{2}}\right)\nonumber\\
		&=\int\;dE\;dE'A^*(E)A(E')e^{-i(E-E')\tau}\int\;dR\;R^{n+2} J_{\nu_a}\left(\lambda_{E'}R^{\frac{3}{2}}\right)J_{\nu_a}\left(\lambda_E R^{\frac{3}{2}}\right)\\
		&=\bar{R}(a,\tau).\nonumber
		\end{align}
		
		The expectation value for this operator is independent of parameter $b$.
		
		\item $\mathbf{\hat{P}^n:}$ Using the representation of momentum operator given in \eqref{mo}, an arbitrary power of momentum operator takes the form,
		\begin{align}
		\hat{P}^n=(-i\hbar)^nR^{-\frac{1}{2}(1-a-2b)}\frac{d^n}{dR^n}R^{\frac{1}{2}(1-a-2b)}.
		\end{align}
		The expectation value of this operator in general wave packet is,
		\begin{align}
		\braket{\Psi|\hat{P}^n|\Psi}&=(-i\hbar)^n\int_{0}^{\infty}dR\;R^{1-a-2b}R^{-\frac{1}{2}(1-a-2b)}\Psi^*\frac{d^n}{dR^n}\left(R^{\frac{1}{2}(1-a-2b)}\Psi\right)\nonumber\\
		&=(-i\hbar)^n\int\;dEdE'A^*(E)A(E')e^{-i(E-E')\tau}\int\;dR\;R^{\frac{1}{2}(1-a-2b)}R^{\frac{1}{2}(1+a+2b)} J_{\nu_a}\left(\lambda_{E'}R^{\frac{3}{2}}\right)\nonumber\\
		&\;\;\;\;\frac{d^n}{dR^n}\left(R^{\frac{1}{2}(1-a-2b)}R^{\frac{1}{2}(1+a+2b)}J_{\nu_a}\left(\lambda_E R^{\frac{3}{2}}\right)\right)\nonumber\\
		&=(-i\hbar)^n\int\!dEdE'A^*(E)A(E')e^{-i(E-E')\tau}\int\!dR\;R\;J_{\nu_a}\!\left(\lambda_{E'}R^{\frac{3}{2}}\right)\frac{d^n}{dR^n}\!\left(\!RJ_{\nu_a}\!\left(\lambda_E R^{\frac{3}{2}}\right)\right)\\
		&=\bar{P}(a,\tau).\nonumber
		\end{align}
		
		This expectation value also is independent of operator ordering parameter $b$.
		
		\item $\mathbf{\hat{O}=\hat{R}^{m}\hat{P}^n\hat{R}^m:}$ The expectation value of this operator in general wave packet is given by,
		\begin{align}
		\braket{\Psi|\hat{O}|\Psi}&=(-i\hbar)^n\int dR\;R^{1-a-2b}R^{m-\frac{1}{2}(1-a-2b)}\Psi^*\frac{d^n}{dR^n}\left(R^{m+\frac{1}{2}(1-a-2b)}\Psi\right)\nonumber\\
		&=(-i\hbar)^n\int\;dEdE'A^*(E)A(E')e^{-i(E-E')\tau}\int\;dR\;R^{m+\frac{1}{2}(1-a-2b)}R^{\frac{1}{2}(1+a+2b)} J_{\nu_a}\left(\lambda_{E'}R^{\frac{3}{2}}\right)\nonumber\\
		&\;\;\;\;\frac{d^n}{dR^n}\left[R^{m+\frac{1}{2}(1-a-2b)}R^{\frac{1}{2}(1+a+2b)}J_{\nu_a}\left(\lambda_E R^{\frac{3}{2}}\right)\right]\nonumber\\
		&\!\!=\!(-i\hbar)^n\!\int\!dEdE'A^*(E)A(E')e^{-i(E-E')\tau}\!\int\!dR\;R^{m+1}\!J_{\nu_a}\!\left(\lambda_{E'}R^{\frac{3}{2}}\!\right)\!\frac{d^n}{dR^n}\!\left[R^{m+1}\!\!\;J_{\nu_a}\!\left(\!\lambda_E R^{\frac{3}{2}}\!\right)\!\right]\\
		&=\bar{O}(a,\tau).\nonumber
		\end{align}
		
		This expectation value also is independent of operator ordering parameter $b$. The second representation we use is Weyl ordering.
		
		\item $\mathbf{\hat{O}=\frac{1}{2}(\hat{R}^{m}\hat{P}^n+\hat{P}^n\hat{R}^m):}$ The expectation value of this operator in a general wave packet is given by,
		\begin{align}
		\braket{\Psi|\hat{O}|\Psi}=&\frac{(-i\hbar)^n}{2}\int dR\;R^{1-a-2b}\Psi^*\biggr[R^{m-\frac{1}{2}(1-a-2b)}\frac{d^n}{dR^n}\left(R^{\frac{1}{2}(1-a-2b)}\Psi\right)+R^{-\frac{1}{2}(1-a-2b)}\nonumber\\
		&\quad\frac{d^n}{dR^n}\left(R^{m+\frac{1}{2}(1-a-2b)}\Psi\right)\biggr]\nonumber\\
		=&\frac{(-i\hbar)^n}{2}\int\;dEdE'A^*(E)A(E')e^{-i(E-E')\tau}\int\;dR\;\biggr[R^{m+\frac{1}{2}(1-a-2b)}R^{\frac{1}{2}(1+a+2b)} J_{\nu_a}\left(\lambda_{E'}R^{\frac{3}{2}}\right)\nonumber\\
		&\quad\frac{d^n}{dR^n}\left(R^{\frac{1}{2}(1-a-2b)}R^{\frac{1}{2}(1+a+2b)}J_{\nu_a}\left(\lambda_E R^{\frac{3}{2}}\right)\right)+R^{\frac{1}{2}(1-a-2b)}R^{\frac{1}{2}(1+a+2b)} J_{\nu_a}\left(\lambda_{E'}R^{\frac{3}{2}}\right)\nonumber\\
		&\quad\frac{d^n}{dR^n}\left(R^{m+\frac{1}{2}(1-a-2b)}R^{\frac{1}{2}(1+a+2b)}J_{\nu_a}\left(\lambda_E R^{\frac{3}{2}}\right)\right)\biggr]\nonumber\\
		=&\frac{(-i\hbar)^n}{2}\int\;dE\;dE'A^*(E)A(E')e^{- i (E-E')\tau}\int\;dR\;\biggr(R^{m+1}\;J_{\nu_a}\left(\lambda_{E'}R^{\frac{3}{2}}\right)\frac{d^n}{dR^n}\left(R\;J_{\nu_a}\left(\lambda_E R^{\frac{3}{2}}\right)\right)\nonumber\\
		&+R\;J_{\nu_a}\left(\lambda_{E'}R^{\frac{3}{2}}\right)\frac{d^n}{dR^n}\left(R^{m+1}\;J_{\nu_a}\left(\lambda_E R^{\frac{3}{2}}\right)\right)\biggr)\\
		=&\bar{O}(a,\tau).\nonumber
		\end{align}
		
	\end{enumerate}
	The expectation value of all these observables is independent of operator ordering parameter $b$. Therefore, $b$ appears in theory as a free parameter as far as the space of initial conditions of the theory is considered, allowing for different valid representations of $\hat{P}$ and $\hat{H}$. However, since these observables are sensitive to the parameter $a$, we intend to construct an observable whose expectation value closely follows this parameter.
	\subsection{Signature of operator ordering parameter}
	
	It is established that parameter $b$ appear as a free parameter in the model. We can still ask if there is any observable signature of the parameter $a$ in the model. In the current analysis, the question is ill posed as we have made the distribution a function of parameter $a$ in the wave packet defined in Eq. \eqref{wp}. Therefore to answer this question, we have to work with the wave packet \eqref{gwp}. In the given form, the closed-form expressions for various integrals are difficult to compute. To tackle this issue, we will take different operator ordering parameters that will simplify the expression while the distribution parameters $\kappa$ and $\lambda$ remain the same throughout the analysis.

	To achieve that, we take two sets of parameter values, (I) $\kappa=4$, $\lambda=1$, $a=5$ and $b=-3\;\&$ (II) $\kappa=4$, $\lambda=1$, $a=11$ and $b=-6$. For these values, we use the properties of  hypergeometric functions \cite{wolfram}
	\begin{align}
	a&=5,\;b=-3\rightarrow\;_1F_1[4,3;x]=e^x\left(\frac{x}{3}+1\right),\\
	a&=11,\;b=-6\rightarrow\;_1F_1[5,5;x]=e^x.
	\end{align}
	The wave packet then takes the form,
	\begin{align}
	\psi_I(R,\tau)&=\frac{16R^3(27-4R^3-54 i \tau)}{243(1-2 i \tau)^5}e^{-\frac{2R^3}{9\left(\frac{1}{2}- i \tau\right)}},\label{wp1}\\
	\psi_{II}(R,\tau)&=\frac{64R^6}{243(1-2 i \tau)^5}e^{-\frac{2R^3}{9\left(\frac{1}{2}- i \tau\right)}}\label{wp2}.
	\end{align}

	Here we can ask if there is an observable signature of parameter $a$ in the expectation value of the areal radius operator. We compute the expectation value and the standard deviation for the areal radius, 
	\begin{figure}[H]
		\begin{tabular}{l l}
			\includegraphics[scale=0.7]{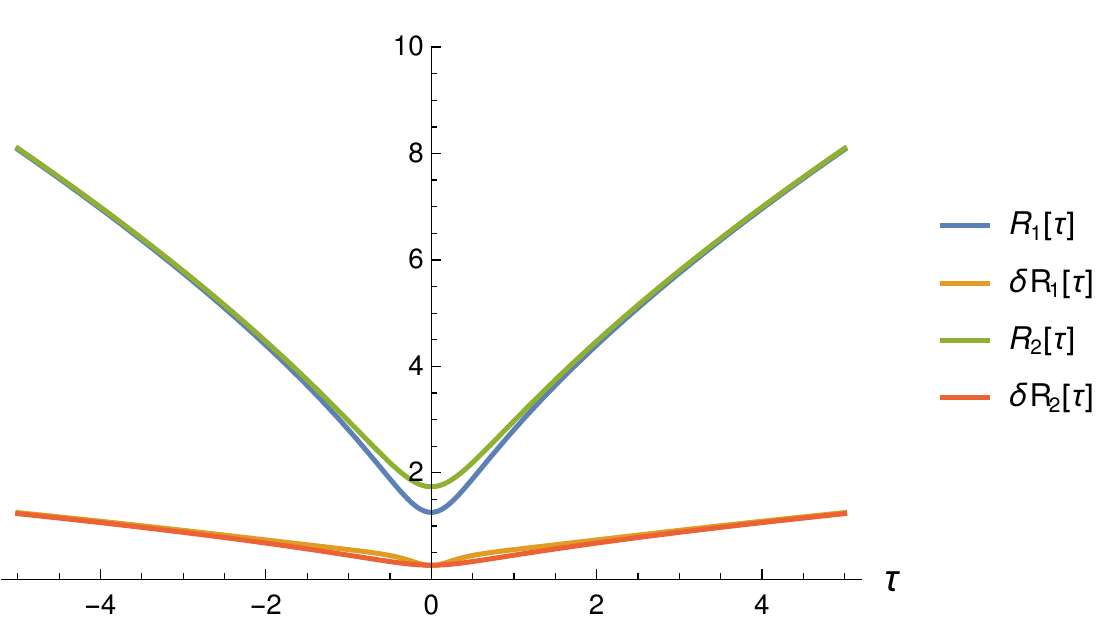} &
			\includegraphics[scale=0.7]{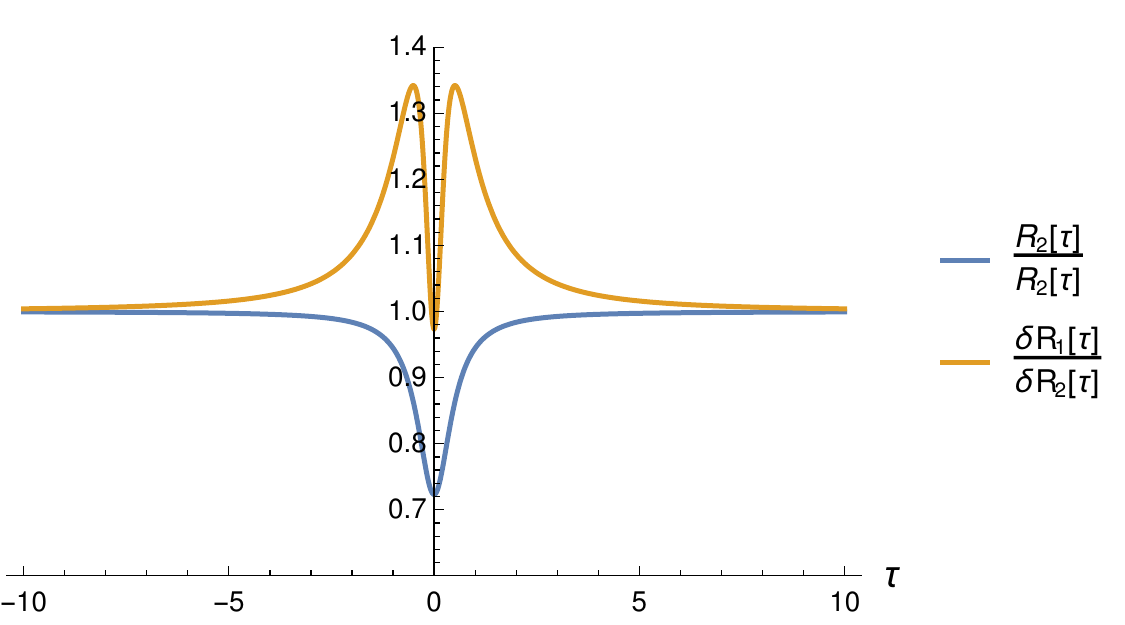}	
		\end{tabular}
		\caption{Expectation value of areal radius and standard deviation for the wave packets 	\eqref{wp1} and \eqref{wp2}. Both expressions match for large $\tau$, and at $\tau=0$ the difference is maximum. The standard deviation of the areal radius also matches for large $\tau$ starts differing as we go toward singularity while it again matches at $\tau=0$.}\label{radius}
	\end{figure}
	\begin{align}
		\bar{R}_I&=\frac{\left(1+4\tau^2\right)^{-\frac{2}{3}}}{240\sqrt[3]{3}}(47+260\tau^2)\Gamma\left[\frac{13}{3}\right],\quad\bar{R}_{II}=\frac{\left(1+4\tau^2\right)^{\frac{1}{3}}}{16\sqrt[3]{3}}\Gamma\left[\frac{16}{3}\right]\\
		\delta R_{I}&=\frac{\sqrt{1200 \left(4 \tau ^2+1\right) \left(308 \tau ^2+41\right) \Gamma \left(\frac{11}{3}\right)-\left(260 \tau ^2+47\right)^2 \Gamma \left(\frac{13}{3}\right)^2}}{240 \sqrt[3]{3} \left(4 \tau ^2+1\right)^{2/3}}\\
		\delta R_{II}&=\frac{\left(1+4\tau^2\right)^{\frac{1}{3}}}{16 \sqrt[3]{3}}\sqrt{24 \Gamma \left(\frac{17}{3}\right)-\Gamma \left(\frac{16}{3}\right)^2}\\
		&\lim_{\tau\rightarrow\pm\infty}\frac{\bar{R}_I}{\bar{R}_{II}}=1,\text{ and}\quad\lim_{\tau \rightarrow 0}\frac{\bar{R}_I}{\bar{R}_{II}}=\frac{47\Gamma\left[\frac{13}{3}\right]}{15\Gamma\left[ \frac{16}{3}\right]}=\frac{47}{65}\approx 0.72 
	\end{align}

	Therefore, very early in the collapse or very late in the expanding phase, the expectation value of the areal radius is insensitive to the operator ordering chosen. On the other hand, the difference between the two is most pronounced at the classical singularity $\tau=0$. It is argued in \cite{kiefer_singularity_2019} that $\bar{R}(0)$ gives the size of the temporary compact remnant of the gravitational collapse, (in the language of \cite{rovelli_planck_2014}) a Planck star. Therefore, the size of the Planck star is sensitive to operator ordering parameter $a$. In the previous section, it is established that the fraction of incoming/outgoing modes of small wave number at $\tau=0$ is sensitive to the size of the dust shell at bounce point (size of the temporary compact remnant).
	Thus it can be inferred that the information about the operator ordering parameter is encoded in the infrared sector of the radiation profile. Whether this feature is an artifact of the minisuperspace reduction or it survives in a more general quantum gravity analysis is a matter of research to be pursued separately. However from Eq. \eqref{smallk}, we can see that although the information regarding the operator ordering ambiguity parameter $a$ is retained in the $k\rightarrow 0$ sector even at late times, the overall amplitude gets progressively diluted. Therefore, the parameter dependence effectively survives only for a finite amount of time, dependent upon the energy profile of the dust cloud \eqref{smallk}. This characteristic is also seen in Fig. \eqref{radius} depicting the areal radius, which also drops the $a$ dependence at late times. This is in tune with the similar characteristics of operator ordering ambiguity parameter dependence reported earlier in literature \cite{Giesel_2006,Bojowald_2014}.
	\section{Conclusions}\label{Consclusions}
	In this work, using the Hermitian extension of the momentum operator on normalizable states, we have studied the mode decomposition of a wave packet constructed for the quantum LTB model,  developed in \cite{kiefer_singularity_2019}. The classical model of such dust shell collapse exhibits black hole singularity, which in the quantum gravity analysis, is replaced by a quantum bounce from a collapsing to an expanding phase. We identify an observable suitable for viable mode decomposition of normalizable physical states, which is the momentum conjugate to the areal radius. Although the momentum operator is not self-adjoint in this setting, one can still obtain its Hermitian extension on the space of normalizable states. 
	Exploiting the freedom offered by this model, we are able to choose the operator ordering parameters $(a,b)$ for which the Hamiltonian operator, as well as the momentum operator, remains Hermitian. This is achieved by first working with $R^2$ measure and choosing the representation which is symmetric with this choice of the inner product. This particular choice puts the constraint $a+2b+1=0$ on operator ordering parameters. We further establish that parameter $b$ appears as a free parameter even with the general measure $R^{1-a-2b}$ in the expectation values of all the phase space observables of the theory, the above constraint does not put any restriction on the space of initial conditions. After identifying incoming and outgoing modes with momentum's eigenstates with positive and negative eigenvalues, we estimate the contribution of incoming/outgoing modes in the contracting/expanding phase.

	We find that at the point of classical singularity $\tau=0$, the number of incoming and outgoing modes become equal for all $k$, representing the quantum bounce. In the preceding contracting branch $\tau <0$, we find that apart from the collapsing dust, we also have a small contribution from outgoing dust modes as well, characterizing emission from the quantum collapse. If we analyze the ratio of incoming to outgoing modes for the contracting branch, it starts from unity at $k=0$ and increases to attain a maximum, and then oscillates back to one. As we keep moving towards the singularity, the contribution of outgoing modes keeps increasing, culminating in the equal number of incoming and outgoing modes at the point of singularity. This behavior is inverted in the expanding branch. After the bounce, outgoing modes start to dominate, with a small fraction of incoming modes present as well. As the dust shell keeps expanding, the fraction of incoming modes keeps decreasing, and the contribution of outgoing modes dominates. 
	
	The infrared segment of the collapsing/expanding branch of the dust collapse contains significant information about the collapse process.  For modes with small wave number, the ratio of incoming to outgoing modes flips as soon as the collapse progresses beyond the bounce $\tau\geq 0$. Furthermore, the contribution of the infrared sector asymptotically provides the bounce radius. Thus not only the infrared sector of the radiation is sensitive to the bounce, it also provides a direct estimator of the bounce radius, providing a unique infrared signature of quantum gravity in the radiation profile of the dust shell. We further observe that for sharper wave packets representing dust clouds with high energy, the contribution of outgoing/incoming modes in the collapsing/expanding phase is smaller when compared with the case of dynamics of low energy dust clouds. Thus low energy dust collapse is more suited for effectively carrying the quantum signatures to the infrared region than a high energetic one.

	Lastly, for the Hermitian extension of momentum operator in a general measure space $R^{1-a-2b}$ compatible with unitary dynamics, we demonstrate that the contributions of incoming (outgoing) radiation in the wave packet follow a similar trend as is observed in $R^2$ measure space. Thus the general behavior in the infrared sector of dust shell in quantum LTB model is unaffected by the choice of the representation of the Hermitian extension of momentum operator and the infrared signatures of the quantum collapse process discussed in this paper, can be considered to be reliably robust. Since the infrared regime carries the imprints of the bounce, the infrared correlators can be expected to be quantum information-wise rich of the quantum gravity imprints too. This line of study, however, will be pursued elsewhere.
	
	
	\section{Acknowledgments}
	HSS would like to acknowledge the financial support from University Grants Commission,
	Government of India, in the form of Junior Research Fellowship (UGC-CSIR JRF/Dec-
	2016/503905). Research of KL is partially supported by the Startup Research Grant of SERB, Government of India (SRG/2019/002202). HSS would like to thank Tim Schmitz for useful correspondences and Ankit Dhanuka for useful discussions.

	%
	\appendix
	
	\section{Orthogonality of Eigenstates of Momentum Operator on \texorpdfstring{$\mathbb{R}^+$}{R+}}\label{or}
	We are interested to see if the eigenstates of momentum operator on the half-line $\mathbb{R}^+$ are orthogonal. Using momentum representation given in \eqref{p} with measure $R^2$, eigenfunctions of momentum are $u_k=e^{ i  kR}/R$. For orthogonality of these states, we need to evaluate the integral,
	\begin{align}
	\braket{u_{k}|u_{k'}}=\int_{0}^{\infty}dR\;R^2 u_{k}^*u_{k'}= \int_{0}^{\infty}dR\; e^{-i (k-k')R}=\delta^{-}(k-k').\label{Inn}
	\end{align}
	The integral in Eq. \eqref{Inn} is the Fourier transform of the Heaviside step function and is given by the Heisenberg distributions \cite{kanwal_generalized_2004},
	\begin{align}
	\delta^{\pm}(k-k')=\int_{0}^{\infty}dR\; e^{\pm i (k-k')R}=\pi\delta(k-k')\pm i\;\mathcal{P}\biggr\{\frac{1}{k-k'}\biggr\}.\label{eq1}
	\end{align}
	The symbol ${\displaystyle \mathcal{P}\biggr\{\frac{1}{x}\biggr\}}$ in the second term is a distribution that takes a function to the Cauchy principle value of the integral ${\displaystyle \int_{-\infty}^{\infty}\frac{f(x)}{x}dx}$. We are interested in the mapping properties of the distribution \eqref{eq1}. To check that, we need to evaluate the integral,
	\begin{align}
	\int_{-\infty}^{\infty}\delta^{\pm}(k-k')f(k)dk=\pi f(k')\pm i\mathcal{P}\biggr\{\int_{-\infty}^{\infty}\frac{f(k)}{k-k'}dk\biggr\}.\label{eq2}
	\end{align}
	Here we will take $f(k)$ as a regular function that decays faster than $k^{-1}$ in the upper half of complex k-plane i.e. satisfies the Jordan's lemma. The Cauchy principal value of the integral is defined as,
	\begin{align}
	\mathcal{P}\biggr\{\int_{-\infty}^{\infty}\frac{f(k)}{k-k'}dk\biggr\}=\lim\limits_{\epsilon\rightarrow 0}\biggr\{\int_{-\infty}^{k'-\epsilon}+\int_{k'+\epsilon}^{\infty}\biggr\}\frac{f(k)}{k-k'}dk.
	\end{align}
	This integral can be computed for given functions $f(k)$ using the residue theorem,
	\begin{align}
	\mathcal{P}\biggr\{\int_{-\infty}^{\infty}\frac{f(k)}{k-k'}dk\biggr\}=i\pi f(k').
	\end{align}
	From Eq. \eqref{eq2}, we get
	\begin{align}
	\int_{-\infty}^{\infty}\delta^{\pm}(k-k')f(k)dk=\pi f(k')\pm i(\pi i f(k'))=\left\{
	\begin{array}{ll}
	0 & \text{for } \delta^{+}(k-k') \\
	2\pi f(k') & \text{for } \delta^{-}(k-k')\\
	\end{array} .
	\right.\label{res1}
	\end{align}
	Using Eq. \eqref{Inn}, we can rewrite Eq. \eqref{res1} as
	\begin{align}
	\int_{-\infty}^{\infty}\braket{u_{k}|u_{k'}}f(k)dk&=2\pi f(k'),\quad\text{ and } \int_{-\infty}^{\infty}\braket{u_{k}|u_{k'}}f(k')dk'=0,\\
	\implies\braket{u_{k}|u_{k'}}&=2\pi\delta(k-k')\theta(k-k')\label{res}.
	\end{align}
	Here $\delta(k-k')$ is the Dirac delta function and $\theta(k-k')$ is the Heaviside theta function. It is apparent that for $k\neq k'$ the inner product vanishes and it has nonzero contribution only at $k=k'$. Thus the distribution \eqref{eq1} has mapping properties akin to a Dirac delta distribution on the space of regular functions, which satisfies Jordan's lemma. The eigenstates of momentum operator can be treated as orthogonal states in this case.\\

	For the case of general measure, the momentum operator is given in \eqref{mo} and the eigenstates of the momentum operator in this case are ${\displaystyle u_k=\frac{e^{ i kR}}{R^{\frac{1}{2}(1-a-2b)}}}$. The inner product of two eigenstates is given by
	\begin{align}
	\braket{u_{k}|u_{k'}}=\int_{0}^{\infty}dR\;R^{1-a-2b} u_{k}^*u_{k'}= \int_{0}^{\infty}dR\; e^{- i(k-k')R}=\frac{1}{2}\delta(k-k')-\frac{ i}{2\pi}\mathcal{P}\biggr\{\frac{1}{k-k'}\biggr\}.
	\end{align}
	The previous results are applicable here as well, and the eigenstates can be treated as orthogonal in this case also.
\end{document}